\documentclass[useAMS,usenatbib]{mn2e}
\usepackage{graphicx,epsfig,amsmath,amssymb,amsbsy,natbib}

\usepackage{graphicx}
\usepackage{subfigure}
\title[Galaxy wide radio induced feedback in the ``Beetle'']{Galaxy wide radio induced feedback in a radio quiet quasar\thanks{Based on observations carried out at the Observatorio Roque de los Muchachos (La Palma, Spain) with OSIRIS on GTC (programmes 
GTC37/14A, GTC1/15A and GTC01-16ADDT) and the Very Large Array (programmes VLA/15A-237 and VLA/16A-088).}}
\author[Villar Mart\'\i n et al.]{M. Villar-Mart\'{i}n$^{1,2}$, B. Emonts$^{1,2}$, A. Cabrera Lavers$^{3,4}$, C. Tadhunter$^5$
\newauthor  D. Mukherjee$^6$, A. Humphrey$^7$, J. Rodr\'\i guez Zaur\'\i n$^5$,   C. Ramos Almeida$^{4,8}$ 
\newauthor   M. P\'erez Torres$^{9,10}$, P. Bessiere$^{11}$ \\
$^1$Centro de Astrobiolog\'{i}a (INTA-CSIC), Carretera de Ajalvir, km 4, 28850 Torrej\'on de Ardoz, Madrid, Spain\\
$^2$Astro-UAM, UAM, Unidad Asociada CSIC, Facultad de Ciencias, Campus de Cantoblanco, E-28049, Madrid, Spain \\
$^3$GRANTECAN, Cuesta de San Jos\'e s/n, E-38712 , Bre\~na Baja, La Palma, Spain \\
$^4$Instituto de Astrof\'\i sica de Canarias, V\'\i a L\'actea s/n, E-38200 La Laguna, Tenerife, Spain \\
$^5$Department of Physics and Astronomy, University of Sheffield, Sheffield, S3 7RH, UK \\
$^6$Research School of Astronomy and Astrophysics, Australian National University, Canberra, ACT 2611, Australia \\
$^7$Instituto de Astrof\'{i}sica e Ci\^encias do Espa\c{c}o, Universidade do Porto, CAUP, Rua das Estrelas, PT4150-762 Porto, Portugal \\
$^8$Departamento de Astrof\'\i sica, Universidad de La Laguna (ULL), E-38205 La Laguna, Tenerife, Spain \\
$^9$Instituto de Astrof'sica de Andaluc'a (IAA-CSIC), Apdo. 3004, 18080 Granada, Spain \\
$^{10}$Departamento de F\'\i sica Te\'orica, Universidad de Zaragoza, C/ Pedro Cerbuna 12, E-50009, Zaragoza, Spain \\
$^{11}$Instituto de Astrof\'\i sica, Facultad de F\'\i sica, Pontificia Universidad Cat\'olica de Chile, Casilla 306, Santiago 22, Chile \\
}

\begin{document}

\date{Accepted ?.
      Received ?;
      in original form ?.}

\pagerange{\pageref{firstpage}--\pageref{lastpage}}
\pubyear{2013}

\maketitle

\label{firstpage}

\begin{abstract}

 We report the discovery of a radio quiet type 2 quasar (SDSS J165315.06+234943.0 nicknamed the ``Beetle'' at $z$=0.103)  with unambiguous  evidence for active galactic nucleus (AGN) radio induced feedback acting across  a total extension of $\sim$46 kpc and  up to  $\sim$26 kpc from the AGN.  To the best of our knowledge, this is the first radio quiet system where radio induced feedback has been securely identified at  $\gg$several kpc from the AGN.
The morphological, ionization and kinematic properties of the extended ionized gas  are correlated with the radio structures. 
 We find along the radio axis (a) enhancement of the  optical line emission at the location of the radio hot spots (b) turbulent gas kinematics (FWHM$\sim$380-470 km s$^{-1}$)  across the entire spatial range circumscribed by them   (c)   ionization minima for the turbulent gas  at the location of the  hot spots, (d)  high  temperature  $T_ {\rm e}\ga$1.9$\times$10$^4$ K at the NE  hot spot. Turbulent gas is also  found far from the radio axis, $\sim$25 kpc in the perpendicular direction.
We propose a scenario in which the  radio structures have perforated the interstellar medium of the galaxy and escaped into the circumgalactic medium.  While advancing, they have interacted with in-situ gas  modifying its properties.   Our results show that jets of modest power can be the dominant feedback mechanism acting across huge volumes in radio quiet systems, including highly accreting luminous AGN, where  radiative mode feedback may be expected.

\end{abstract}


\section{Introduction}
\label{Sec:intro}

 Feedback induced by the activity of supermassive black holes (SMBH) in massive galaxies is thought to play a critical role in their evolution by means of regulating the amount of gas available for star formation and black hole growth. 
This type of AGN (active galactic nuclei) feedback refers to the processes of interaction
between the energy and radiation generated by accretion onto the massive black hole and the gas in the host galaxy (Fabian \citeyear{fab12}).  This may give answers to some fundamental questions in cosmology (Silk \& Rees \citeyear{silk98},  King \citeyear{king03}, Silk \& Mamon \citeyear{silk12}), such as the discrepancy between the predicted and observed high mass end of the galaxy luminosity function and the observed correlations  between the black hole mass and some  properties of the spheroidal component in galaxies (Ferrarese \& Merritt \citeyear{fer00}, McConnell \& Ma \citeyear{mcc13}). Hydrodynamic simulations show that the energy output of these outflows   can indeed regulate the growth and activity of black holes and their host galaxies (di Matteo et al. \citeyear{dim05}). For this,  the intense flux of photons and particles produced by the AGN must sweep the galaxy clean of interstellar gas and terminate star formation and the activity of the SMBH (Fabian \citeyear{fab12}). The most powerful outflows with potentially the most extreme effects on the environment are expected in quasars, the most powerful AGN (Page et al. \citeyear{page12}, Woo et al. \citeyear{woo17}). Observational evidence for such dramatic an impact   is  however controversial.

Since their recent discovery, optically selected type 2  quasars (QSO2,  Zakamska et al. \citeyear{zak03}, Reyes et al. \citeyear{rey08}) have been unique systems for investigating  the way  feedback works in the most powerful AGN.  
The active  nucleus is  obscured and this 
allows a detailed study of many properties  of the surrounding medium without the overwhelming glare of the quasar, which strongly affects related   studies
of their unobscured type 1  counterparts (QSO1).

  During recent years it has become clear that ionized outflows are  ubiquitous  in QSO2 at different $z$ (Villar Mart\'\i n et al. \citeyear{vm11}, \citeyear{vm16},   Greene et al. \citeyear{gre11},  Harrison et al. \citeyear{har12}, Liu et al. \citeyear{liu13},  Harrison et al.  \citeyear{har14},  Forster-Schreiber et al. \citeyear{for14}, McElroy et al. \citeyear{mce15}). Extreme motions are often measured, with FWHM$>$1000 km s$^{-1}$ and typical velocity shifts $V_S\sim$several$\times$100 km s$^{-1}$. 
The outflows are triggered by AGN related processes (e.g. Villar Mart\'\i n et al.  2014, hereafter \cite{vm14}, Zakamska \& Greene \citeyear{zak14}).  

 The case for effective gas ejection and star formation truncation  exerted by the ionized outflows is less clear. 
Recent integral field spectroscopic studies of  QSO2 at $z$$\la$0.7 have suggested that large scale, wide angle AGN driven ionized  outflows are prevalent in these systems and their action  can be exerted across many kpc,
even the entire galaxy (Liu et al. \citeyear{liu13},  Harrison et al.  \citeyear{har14},   McElroy et al. \citeyear{mce15}). The  kinetic energies, $\dot E_{\rm o}$, outflow
masses, $M_{\rm o}$, and mass outflow rates, $\dot M_{\rm o}$, are estimated to be large enough  for a significant impact on their
hosts. In the proposed scenario, processes related to the accretion disk  (thermal, radiation and magnetic driving) launch a
relativistic wide angle wind  which then shocks
 the surrounding gas and drives the outflow  (e.g. Proga \citeyear{prog07}, Zubovas \& King \citeyear{zub14}).
  On the other hand, the above observational  results have been questioned by different authors (\cite{vm14}, Karouzos et al. \citeyear{kar16}, Villar Mart\'\i n et al. \citeyear{vm16}, Husemann et al. \citeyear{hus16}; see also Husemann et al.  \citeyear{hus13} for QSO1) who find that the ionized outflows are  constrained in general within $R_{\rm o}$$\la$1-2 kpc.  This
raises doubts about their impact on the host galaxies, since their effectiveness as a feedback mechanism would be significantly reduced.

Another mechanism of AGN feedback is that induced by relativistic jets. The jets originate in the vicinity of the SMBH, on scales of a few to 100 times the gravitational radius. Three dimensional relativistic hydrodynamic simulations of interactions of AGN jets with a dense turbulent two-phase interstellar medium, show that 
 the originally well-directed jets form  an energy-driven almost  spherical  bubble that can affect the gas to distances up to several kpc from the injection region, depending on the jet power. The shocks resulting from such interactions create a multiphase ISM and radial outflows (e.g. Mukeherjee et al. \citeyear{muk16}). 
 
 Observational studies of powerful radio galaxies and radio loud QSO1  at different $z$ show that the interaction between the radio structures and the ambient gas can  indeed induce outflows that can extend across many kpc,  sometimes well outside the galaxy boundaries (Tadhunter et al. \citeyear{tad94}, McCarthy et al. \citeyear{mcc96}, Villar Mart\'\i n et al. \citeyear{vm99},\citeyear{vm03}, Sol\'orzano I\~narrea et al. \citeyear{sol01}, Humphrey et al. \citeyear{hum06}, Fu \& Stockton \citeyear{fu09}).  They  may have enormous energies sufficient to eject a large fraction of the gaseous content out of the galaxy and/or quench star formation, thus having an important impact on the evolution of the host galaxies  (Morganti et al. \citeyear{mor05}, Nesvadba et al. \citeyear{nes06}, \citeyear{nes08}).

 This feedback mechanism (hereafter radio induced feedback)  has been proposed to explain the gas cooling problem at the centre of galaxy clusters (see Fabian \citeyear{fab12} for a review). It  may also prevent the formation of extremely bright galaxies at the centre of clusters (e.g.  Sijacki \& Springel \citeyear{sij06}). Such scenarios involve a central dominant elliptical of a cluster going through a powerful radio loud phase. Simulations by \cite{gas12}  show that radio induced feedback can have significant effects  also in more modest environments, including isolated ellipticals.  It could explain the evidence for reduced cooling also in elliptical galaxies
 and the related absence of cold gas in these systems (Mathew \& Baker \citeyear{mat71}, Best et al. \citeyear{best06}).

The role of radio induced feedback has not been sufficiently explored in  radio quiet quasars. Only $\sim$15$\pm$5\%   of QSO2 
 are thought to be  radio loud  (Lal \& Ho \citeyear{lal10}). The same percentage applies to QSO1 (Kellermann et al. \citeyear{kel94}). However, this does not discard radio induced outflows  as a potentially significant feedback mechanism.  Radio-gas interactions\footnote {By ``radio-gas interactions'' we will refer to the interactions  between any AGN related radio structure (jets, lobes, hot spots) and the ambient gas} and the associated outflows have been observed in radio quiet AGN for several decades (e.g. Wilson \& Ulvestad \citeyear{wil83}, Whittle \citeyear{whi85},  \citeyear{whi92}). The most extreme  ionized outflows are often found in
objects (including  QSO2) with some degree of radio activity, even if they are not radio loud (e.g.   Mullaney et al. \citeyear{mul13}, \cite{vm14},   Husemann et al. \citeyear{hus13}, \citeyear{hus16}, Zakamska \& Greene \citeyear{zak14}). Its  significant role as a feedback mechanism has been quantified in  several low luminosity AGN (Seyfert galaxies) (Tadhunter et al. \citeyear{tadh14}, Alatalo et al. \citeyear{ala15}). Although the effects are obvious in a small region ($R\la$1 kpc from the SMBH), the  outflows can disturb and heat the molecular gas, thus having the potential to  quench star formation (Guillard et al. \citeyear{gui12}).

Large scale effects ($\gg$ several kpc) due to radio induced feedback  in radio quiet AGN are usually not expected, nor are they usually observed. 
Traditionally, it is assumed that radio quiet AGN contain low power jets which will decelerate and lose their power after a few kpc. However, different results suggest that radio induced feedback may indeed occur across many kpc in some systems. On one hand, radio-gas interaction simulations show that while a  high power jet ($P_{\rm jet}\ga$10$^{45}$erg s$^{-1}$) can remain  relativistic to large scales and drill  with relative ease through tens of kpc,   a jet with low mechanical power ($P_{\rm jet}\la$10$^{43}$erg s$^{-1}$), although less efficient in accelerating clouds, is trapped in the interstellar medium (ISM) for a longer time and hence can affect the ISM over a larger volume (Bicknell \citeyear{bic94}, Mukherjee et al. \citeyear{muk16}).     On the other hand,  some radio quiet quasars  associated with modest or low power jets show large radio sources with sizes in the range of tens or even hundreds of kpc (Kellerman et al. \citeyear{kel94}). Thus, it is clear that  these radio sources can  escape the galaxy boundaries and expand into the circumgalactic medium (CGM).\footnote{Following Tumlinson et al. \citeyear{tum17}, we consider the CGM as the gas surrounding galaxies outside their disks or ISM and inside their virial radii.}

However, to the best of our knowledge, the effects of radio-gas interactions have not been directly observed  beyond scales of $\sim$several kpc from the AGN in such systems. The ``Teacup''  QSO2 ($z=$0.085), which   shows    $\sim$10-12 kpc radio/optical bubbles, may be the radio quiet AGN where the effects of   radio-gas interactions have been identified across  the largest spatial scale so far (Harrison et al. \citeyear{har15}, Ramos Almeida et al. \citeyear{ram17}).  It is not clear, however,  whether the bubbles  have been inflated  instead by a wide angle large scale AGN driven wind (Harrison et al. \citeyear{har15}).

We report here unambiguous evidence for radio induced feedback working across a total extension of $\sim$46 kpc and up to $\sim$26 kpc from the AGN, well into the CGM, in a radio quiet QSO2.

We present a detailed study of  SDSS J165315.06+234943.0 (nicknamed the ``Beetle''   hereafter) at $z$=0.103 
 based on GTC optical images and  long slit spectroscopy and VLA radio maps. The data reveal that  radio induced feedback  is modifying the morphological, kinematic and physical properties of the gaseous environment on scales from (possibly) $\sim$1 kpc up to the CGM.

Previous knowledge on the ``Beetle''  is summarized in Sect. \ref{Sec:object}. The  observations,  reduction and analysis methods of the different data sets (optical imaging and spectroscopy, radio maps) are described in  Sect. 
\ref{Sec:observations}.  Analysis and results are presented in  Sect. \ref{Sec:results} and discussed in  Sect. \ref{Sec:discussion}. The summary and conclusions are in   Sect. \ref{Sec:conclusions}.

We adopt $H_{0}$=71 km s$^{-1}$ Mpc$^{-1}$, $\Omega_{\Lambda}$=0.73 and
$\Omega_{m}$=0.27.  This gives an
arcsec to kpc conversion of 1.87 kpc arcsec$^{-1}$ at $z$=0.103. 

\subsection{The radio quiet QSO2 SDSS J165315.06+234943.0}
\label{Sec:object}

The ``Beetle''      was originally selected by    \cite{rey08} for their Sloan Digital Sky Survey (SDSS, York et al. \citeyear{york00}) catalogue of optically
selected QSO2 at $z\la$0.8. Its [OIII]$\lambda$5007 luminosity, log($L_{\rm [OIII]}$)=42.6,  corresponds to a bolometric luminosity log($L_{\rm bol}$)$\sim$46.0 (Stern \& Laor \citeyear{ste12}), in the range of quasars (Woo \& Urry \citeyear{woo02}). 

The information and calculations that follow are provided because they will be relevant in later sections.
\begin{itemize}

\item Radio loudness.

The ``Beetle''  has been detected at 1.4 GHz in the VLA-FIRST survey with a peak flux of 6.4 mJy, while its peak flux  in the VLA NVSS survey  is 9.4 mJy.        
  In order to classify this QSO2 according to the radio-loudness, we have used the rest-frame radio power at 5 GHz $P_{\rm 5GHz}$
and log($L_{\rm [OIII]}$) (see Fig. 7 in Lal \& Ho \citeyear{lal10}; see also Xu et al. \citeyear{xu99}). These two luminosities separate objects clearly into 
the two families of radio loud and radio quiet AGN with a significant gap between them (Fig. \ref{fig:rlrq}). We calculate $P_{\rm 5GHz} = 4 \pi D_{\rm L}^{2} S_{\rm 5GHz} (1+z)^{-1-\alpha}$ where  $D_{\rm L}$ is the luminosity distance, $S_{\rm 5GHz}$ is the observed   flux density at 5 GHz and $\alpha$ is the spectral index such that $S_{\nu} \propto \nu^{\alpha}$. 
We compute  $S_{\rm 5GHz}$ from the available $S_{\rm 1.4GHz}$. 
 The index $\alpha$ is unknown. For  $\alpha =+0.094$ (Lal \& Ho \citeyear{lal10}), log$(P_{\rm 5GHz})=$30.4 (erg s$^{-1}$ Hz$^{-1}$). Considering the range -1.1$\le \alpha\le$+0.85  spanned by QSO2s (Lal \& Ho \citeyear{lal10}), we obtain log(P$_ {\rm 5GHz}$)=30.4$^{+0.4}_{-0.6}$ erg s$^{-1}$ Hz$^{-1}$.   Fig. \ref{fig:rlrq} shows that the ``Beetle'' is a radio quiet QSO2, and this conclusion is not affected by the uncertainty in $\alpha$.

\begin{figure}
\includegraphics{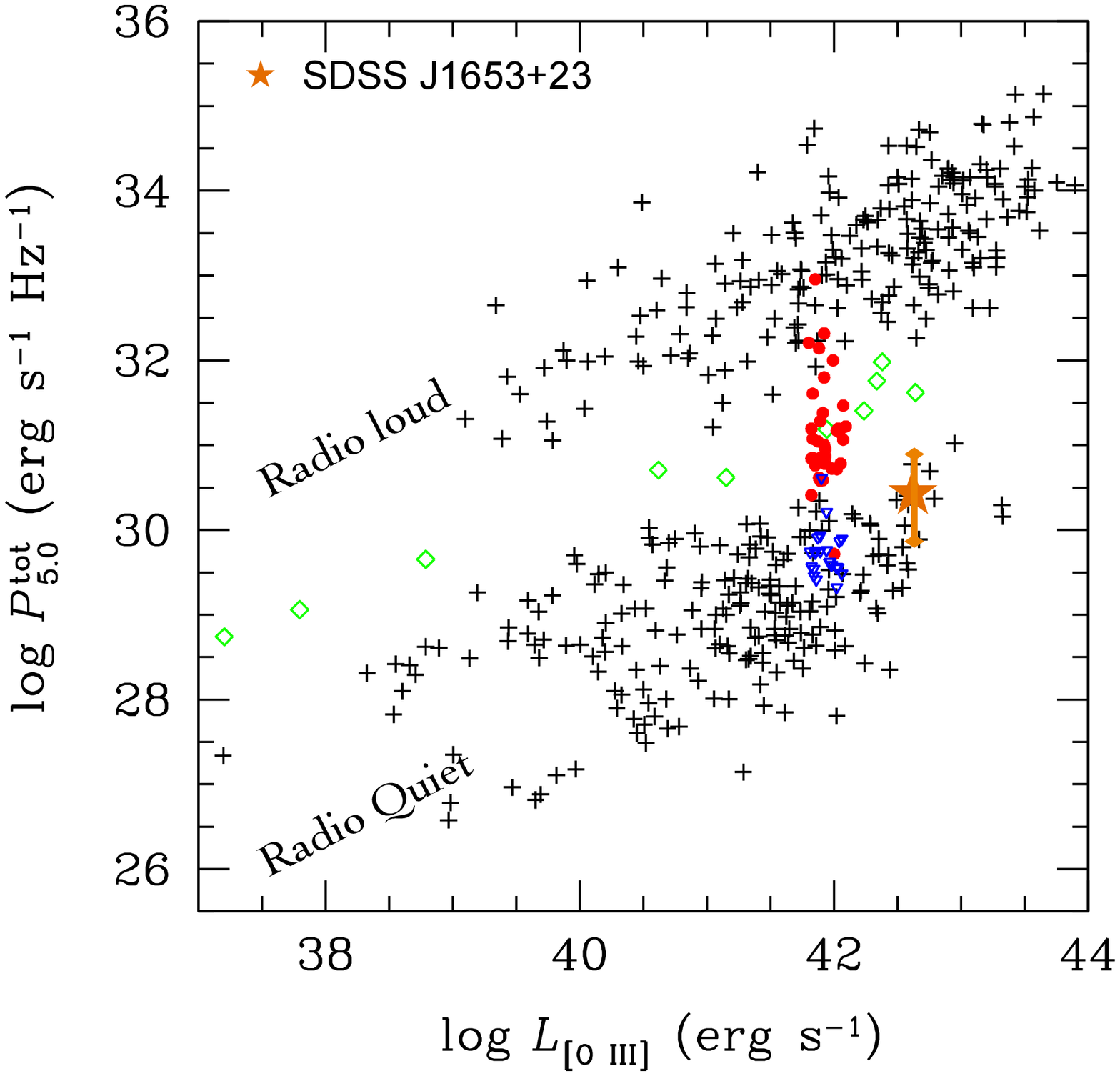}
\vspace{3.5in}
\caption{Classification of the ``Beetle" according to its radio loudness. The figure is the  same as Fig. 7 in Lal \& Ho (2010) with the location of our object shown as a beige star. The error-bar accounts for the uncertainty in the radio spectral index $\alpha$. The plus symbols come from the AGN sample of Xu et al. (1999), with the open green diamonds representing their  radio intermediate sources. The Lal \& Ho (2010) sample of QSO2 at 0.3$\la z \la$0.7 
are represented by filled red circles, with upper limits marked with open blue triangles. The ``Beetle''  QSO2 is radio quiet.}             
\label{fig:rlrq}
\end{figure}

\item Infrared luminosity and its origin.

The IRAS fluxes for the ``Beetle" are $S_ {\rm 12\mu m}$=0.10, $S_ {\rm 25\mu m}$=0.15, $S_ {\rm 60\mu m}$=0.49 and $S_ {\rm 100\mu m}\la$1.01 Jy. Following \cite{san96}, the  total infrared luminosity is 4.3$\times$10$^{11}$ L$_{\rm \odot} < L_ {\rm IR(8-1000 \mu m)} \la$ 5.5$\times$10$^{11}$ L$_{\rm \odot}$\footnote{$k$ correction is negligible for z=0.103}  implying that the object is  in the regime of luminous infrared galaxies (LIRGs, 10$^{11}$$\le L_{\rm IR}/L_{\rm \odot}<$10$^{12}$). Assuming  $S_ {\rm 60\mu m}/S_ {\rm 100\mu m}\ge$1.4 as observed in LIRGs (e.g. Lu et al. \citeyear{lu14}, Pearson et al. \citeyear{pea16}) then $L_ {\rm IR(8-1000 \mu m)}$=(5.0$\pm$0.4)$\times$10$^{11}$ L$_{\rm \odot}$. 
The far infrared luminosity is $L_ {\rm FIR(40-500 \mu m)}$=(1.7$\pm$0.3)$\times$10$^{11}$ L$_{\rm \odot}$.

If $L_{\rm IR}$ were to dominated by a starburst contribution, this  would imply a star forming rate SFR=87$\pm$7 M$_ {\odot}$ yr$^{-1}$ (Kennicutt \citeyear{ken98}).  However, this is a gross upper limit since    $L_{\rm IR}$
 is  contaminated (maybe dominated) by AGN heated dust. This is based on two arguments. Following \cite{hel85}, we have calculated $q = \rm log [\frac{F_{\rm FIR}/3.75 \times 10^{12} \rm Hz}{S_{\rm 1.4 GHz}}]$=1.87$\pm$0.09, where $F_ {\rm FIR}$ is the far infrared flux in erg s$^{-1}$ cm$^{-2}$ and $S_{\rm 1.4 GHz}$ is in units of  erg s$^{-1}$ cm$^{-2}$ Hz$^{-1}$. For the ``Beetle" $q$ is within the range measured for radio emitting AGN ($<$$q_{\rm AGN}$$>$=2.0 with rms scatter $\sigma_ {\rm AGN}$=0.59) and outside the range measured for star forming galaxies ($<$$q_{\rm SF}$$>$=2.3 with rms scatter $\sigma_ {\rm AGN}$=0.18).  The IR colour $\alpha_ {\rm IR}$=log($S_{\rm 25 \mu m}/S_{\rm 60 \mu m}$)/log(60/25)=-1.3  is also  consistent with AGN values ($\alpha_ {\rm IR}>$-1.5, in comparison with  $\alpha_ {\rm IR}<$-1.5 for  star forming galaxies, Mauch \& Sadler \citeyear{mau07}).

\item Stellar velocity dispersion and stellar mass

Two components of the MgI triplet at  5169  \AA\ and 5185 \AA\  are clearly detected in the nuclear GTC spectrum. They have $\sigma$=210$\pm$25 km s$^{-1}$ and 220$\pm$33 km s$^{-1}$ respectively, which is  in reasonable agreement with
$\sigma_ {*}$=169$\pm$8 km s$^{-1}$ provided by the the SDSS Data Release 10 (DR10)   kinematic fits.
We will assume $\sigma_*\sim$200 km s$^{-1}$.

Estimated stellar masses for the ``Beetle''  are in the range 10.64$\la$log$(\frac{M_{\rm *}}{M_{\rm \odot}}$)$\la$11.34, with a median value 11.2. The different values have been retrieved from the SDSS DR10 and the Vizier catalogue  (Ochsenbein al. \citeyear{och00})  based on \cite{men14}.  According to these authors (see also  Simmard et al. \citeyear{sim11}), the spheroidal component contains a mass  log($M_{\rm sph}/M_{\rm \odot}$)=11.0.

\item Black hole mass and Eddington luminosity.

	We have constrained the black hole mass $M_ {\rm BH}$ using the correlation between stellar velocity dispersion $\sigma_*$ and $M_ {\rm BH}$  for early type galaxies  (McConnell \& Ma \citeyear{mcc13}): log($M_ {\rm BH}/M_{\rm \odot}$)= 8.39+5.20 log($\sigma_*$/200 km s$^{-1}$)$=$8.4, where $\sigma_*$$\sim$200 km s$^{-1}$. This is consistent with the value log($M_ {\rm BH}/M_{\rm \odot}$)= 8.5 obtained from the $M_ {\rm BH}$ vs. bulge (or spheroidal component) mass  $M_{\rm sph}$ correlation log($M_{\rm BH}/M_{\rm \odot}$)=8.46+1.05 log($M_{\rm sph}/10^{11} M_{\rm \odot}$). The implied Eddington luminosity  is $L_{\rm Edd}$ =1.26 $\times$10$^{38}$ $(M_ {\rm BH}/M_{\rm \odot}$) erg s$^{-1}\sim$3.2 $\times$ 10$^{46}$ erg s$^{-1}$.

\item A nuclear ionized outflow driven by the nuclear activity was identified in the SDSS  optical spectrum of this QSO (\cite{vm14}),
 with  FWHM$\sim$970 km s$^{-1}$ and a velocity blueshift $V_{\rm s}\sim$ -90 km s$^{-1}$. 

\end{itemize}

\section{Observations, data reduction and analysis}
\label{Sec:observations}

\begin{figure*}
\includegraphics{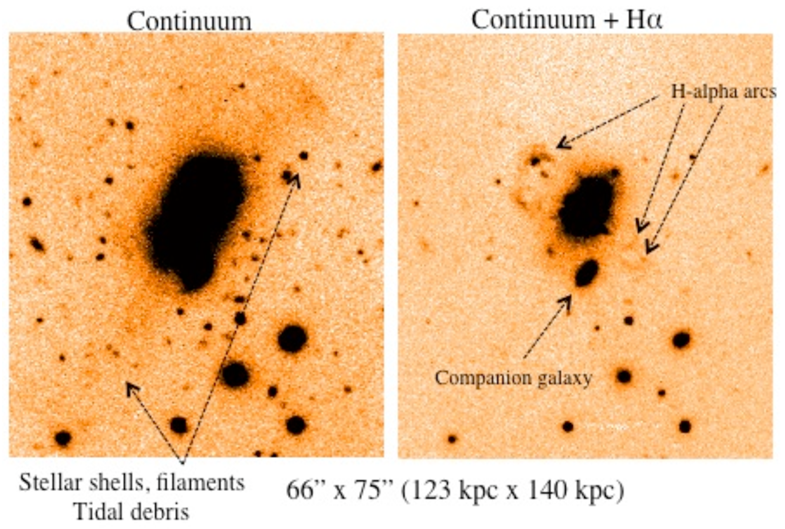}
\vspace{3.7in}
\caption{Left: optical GTC-Osiris  continuum (left) and narrow band H$\alpha$+continuum (right) images of the ``Beetle''. Both images show the same field of view. The continuum image shows prominent tidal features (shells, filaments) well outside the main body of the galaxy. They are not detected in the narrow band  image. This shows instead an intricate set of knots/arcs, shaping structures reminiscent of optical bubbles or bow shocks.  They run roughly perpendicular to the main galaxy axis and the axis of the tidal debris identified in the continuum image. The filaments are not detected in the continuum image. They consist of ionized gas. }             
\label{fig:gtc}
\end{figure*}

\subsection{GTC imaging}
\label{Sec:gtcim}

\subsubsection{Observations}

  H$\alpha$ Tunable Filter
(TF) imaging observations were taken in service mode on July 25th 2014 (programme  GTC37-14A) for the ``Beetle'' using the optical imager and long slit spectrograph OSIRIS\footnote{http://www.gtc.iac.es/en/pages/instrumentation/osiris.php}
mounted on the 10.4m Gran Telescopio Canarias (GTC).  OSIRIS consists of a mosaic of two 2048
$\times$ 4096 Marconi CCD42-82 (with a 9.4 arcsec gap between them) and covers the
wavelength range from 0.365 to 1.05 $\mu$m with a field of view of 7.8 x 7.8
arcmin and a pixel size of 0.127 arcsec. However, the OSIRIS standard
observation modes use 2$\times$2 binning, hence the effective pixel size during
our observations was 0.254 arcsec.

Our goal is to detect extended ionized gas (H$\alpha$) associated with the QSO2. For this, 
a filter FWHM of 20 \AA\ was used ($\sim$830 km s$^{-1}$ at the QSO2 $z$) centred on the redshifted H$\alpha$ and taking into account the dependence of the wavelength  observed with the red TF with distance relative to the optical centre. The order sorter filter f723/45 was also used. The FWHM is   adequate to cover the velocity range expected to be spanned by any plausible
extended ionized features.   Contamination by [NII]$\lambda\lambda$6548,6583 is expected to
be negligible, since the lines are located at  -16.5 \AA\ and +22 \AA\  relative to H$\alpha$, and H$\alpha$ is the strongest of the three lines at all locations (our long spectroscopy confirms this). The total exposure time on source was 2700 seconds distributed in 3$\times$900 second exposures. To be able to
correct for ghost images and cosmic rays we dithered between three positions,
moving the telescope $\pm$15 arcsec in RA and Dec.

To sample the continuum near the H$\alpha$+[NII]
complex we took one continuum image to the blue of H$\alpha$ using
directly the OSIRIS Red Order Sorter (OS) Filter f666/36, which covers the 	6490 - 6845 \AA\ spectral window
(5884 - 6206 \AA\ rest frame). This filter was used to avoid emission line contamination (by, for instance, [OI]$\lambda$6300). The disadvantage is that objects
with a steep continuum slope (for instance, very red objects) will leave prominent  residuals due to the significant shift in $\lambda$ relative to H$\alpha$. Thanks to the complementary spectroscopic data, we can confirm that this has no impact on our analysis and interpretation.
The total exposure time of the continuum image was 600 seconds split into 3$\times$200 second exposure. The same dithering pattern was applied as for the H$\alpha$ image.

\subsubsection{Reduction process}

 The TF and the OS filter OSIRIS images were bias and
flat-field corrected as usual, using a set of bias frames and dome
flats. 

We observed a photometric  standard star for flux calibration using the exact same set up as the one
used for the science object. After correcting the  corresponding frames for bias and flat-field we applied aperture
photometry using the routine PHOT in {\it pyIRAF}. The PHOT output gives us the
integrated number of counts within the aperture for the standard star
observed. We then used a customized Python routine that generates a TF
transmission curve for a given central wavelength and FWHM. The resulting TF transmission
curve was normalized and multiplied by the spectra of the standard star in
units of erg s$^{-1}$ cm$^{-2}$ \AA$^{-1}$. This gives us the expected flux
of the standard star in the corresponding TF and OS filter in units of erg
s$^{-1}$ cm$^{-2}$. We then multiply these numbers by the exposure time used for
the standard star observations and divide the result by the output from
PHOT. The result will be a flux calibration factor ($F$) in units of erg
cm$^{-2}$ counts$^{-1}$. Finally, to calibrate in flux our science frames we
divide them by their corresponding exposure times and multiply them by $F$,
which transforms the units of the frames after bias and flat field correction
(counts) into erg s$^{-1}$ cm$^{-2}$.

Once the images are calibrated in flux, we need to subtract the sky emission. When using a TF imaging
technique the wavelength observed changes relative to the optical centre. In the
case of OSIRIS, this change is given by  \cite{gon14}:

\begin{equation} 
{\lambda} = {\lambda_{0}}-5.04*r^{2}
\end{equation}

Where $\lambda_{0}$ is the wavelength in \AA\  at the optical centre, located at pixel
(2098, 1952) of CCD1 in the case of OSIRIS, and r is the distance from that
centre in arcmin. Therefore, the different sky lines around the wavelength range
selected for the observations are observed in the form of rings of emission
spread over the CCDs. This effect becomes important if the sample objects are
relatively extended and  a careful sky
emission subtraction must be performed.

With this aim we first cut a rectangular region of the CCD that is large enougt
to contain a sizeable region of the sky, and small enough so that the effects of
the sky emission rings are reduced to the minimum possible. Then we used FIT1D in
{\it pyIRAF} to fit a 1-dimensional polynomial function to the overall structure in the x
and y direction respectively. The first x/y 1-dimensional fit is performed only on sky
regions free from emission from stars in the frame and the target observation,
and it is then extended through the entire rectangular cut. The following y/x is
applied to all the rectangular cut at once. The result of this process is a
synthetic image containing only the sky emission. Then this image is directly
subtracted from the original flux calibrated image of the source resulting in an image cleaned from any background and sky
emission.

The next step during the reduction process is subtracting the continuum from the
emission line image. Prior to this they were geometrically aligned. Based on  the centroid of the stars in the aligned images and
 the position of some important features in the galaxy, we estimate that
the alignment accuracy is better than 0.5 pixels. The continuum
emission was then subtracted from the emission line image in order to produce the final
``pure'' H$\alpha$ image.

\subsection{GTC long slit spectroscopy}
\label{Sec:gtcspec}

Long slit spectroscopic observations were performed in service mode with  OSIRIS on GTC between June 26th and July 23rd 2015 (programme GTC1-15A). The   R2500V and R2500R volume-phased holographic gratings (VPHs) were used, that provide a spectral coverage of 4500  - 6000 \AA\ and 5575 - 7685 \AA, respectively, with   dispersions of 0.8 and 1.04 \AA/pixel. A 1.23 arcsec slit width was used at three different orientations (position angle PA 0$^\circ$, PA 48$^\circ$ and PA 83$^\circ$)\footnote{All position angles throughout the text are quoted N to E} in order to map the emission around the quasar. The spectral resolution FWHM$_ {\rm inst}$ measured from several prominent sky lines is 3.37$\pm$0.13 and 4.41$\pm$0.06  \AA\ for R2500V and R2500R respectively. Several 300 second spectra were obtained at each orientation (Table \ref{tab:longslit}), with shifts in the slit direction of 20" between consecutive exposures for both a better fringing correction and a more accurate background subtraction.

Additionally, a series of 3 x 900 second R2500V spectra were taken on 07/03/2016 under GTC DirectorÕs Discretionary Time (programme GTC01-16ADDT).
The same slit width was used  with  PA 140$^{\circ}$. Those spectra where obtained avoiding the quasar nucleus and roughly perpendicular to the main radio axis (PA 48$^\circ$) .

\begin{table*}
\centering
\caption{Log of the GTC Osiris long slit spectroscopic observations} 
\begin{tabular}{cccccc}
\hline
Obs. date &  PA ($^{\circ}$) &  Exposure  & Exposure & Airmass & Seeing \\ 
&  (N to E)  &  R2500V  & R2500R &  \\ \hline
26/27-06-2015 & 48 & 18 x 300 s   &  --  & 1.4  & 0.9" \\
17/07-2015 &  48 & -- & 16 x 300 s &  1.2  & 0.9" \\
18/07-2015 &   0 & 8 x 300 s & 8 x 300 s & 1.1 & 1.1" \\
23/07/2015 &  83 & 8 x 300 s & 4 x 300 s & 1.1 & 1.2" \\ \hline
07/03/2016 &  140 & 3 x 900 s & -- & 1.1 & 1.0" \\ 
 \hline
\end{tabular}
\label{tab:longslit}
\end{table*}

\subsubsection{Reduction process and spectral fitting}

The reduction of the spectra was done using standard procedures and IRAF tasks. Images were first bias and flat-field corrected, by using lamp flats. The 2-dimensional spectra were wavelength calibrated using Xe+Ne+HgAr lamps, with a resulting error consistent with the nominal spectral resolutions of the VPHs. After the wavelength calibration, sky background was subtracted and a two dimensional spectrum was obtained. Flux calibration was done using observations of spectrophotometric standard stars, white dwarfs, obtained the same nights as the scientific spectra. Finally, the diffferent individual spectra were averaged, obtaining a mean spectrum for each of the orientations.

Like the images, the spectra were corrected for atmospheric and foreground galactic extinction ($A_{\rm V}$=0.153).

The spectral profiles of the nuclear emission lines are complex (several kinematic components). They were fitted following the multi-Gaussian procedure described in detail in \cite{vm16}. In order to have a more realistic estimation of the uncertainties, whenever possible, the lines were fitted in two different ways: a) with kinematic constraints based on the results of the [OIII]$\lambda\lambda$4949,5007  fits and b) without prior kinematic constraints. When both methods provided fits of similar quality, the fitted parameter  values (fluxes, FWHM and $V_{\rm s}$) are the average  of all fits. The error for each parameter is chosen as the largest value between the fit errors and the standard deviation of all valid fits (see Villar Mart\'\i n et al. \citeyear{vm16} for more details). The FWHM values were corrected for instrumental broadening by subtracting the instrumental profile FWHM$_{\rm inst}$ in quadrature.

When physically meaningful or mathematically valid fits (e.g. for faint emission lines) could not be obtained without kinematic constraints, the [OIII] doublet lines were used as reference to force the individual kinematic components to have the same velocity shift and/or FWHM.

\subsection{VLA radio maps}
\label{Sec:vla}
Observations with the Karl G. Jansky Very Large Array (VLA) were performed on 13 and 18 July 2015 with the A-configuration (programme 15A-237) and on 01 June 2016 with the B-configuration (programme 16A-088). The on-source exposure time in A-configuration was 0.7\,h and in the B-configuration 3\,h. 

The correlator was set to produce a 1 GHz bandwidth, ranging from 1\,$-$\,2 GHz, with 1 MHz channels. 
We used 3C\,286 for flux calibration, and obtained a 5 min scan on the strong ($\sim$5 Jy) source J1609+2641 every $\sim$50 min for both phase and bandpass calibration. Flagging of radio-frequency interference (RFI), which affected $\sim$40$\%$ of our data, as well as a standard bandpass, phase and flux calibration was performed using the VLA data reduction pipeline in the Common Astronomy Software Applications (CASA) v. 4.6 (McMullin et al. \citeyear{mcm07}). We manually inspected the pipeline-calibrated data to ensure that the quality of the data products was sufficient for further reduction. We subsequently applied one self-calibration to the A-configuration data and two iterative self-calibrations to the B-configuration data to improve the phases. 

The data were imaged by applying a Fourier transform using a multi-frequency synthesis technique, and subsequently cleaning the continuum signal. We imaged the A-configuration data using uniform weighting, which resulted in a beam of 0.80\,$\times$\,0.74 arcsec (PA\,=\,41.8$^{\circ}$) and rms noise level of 0.066 mJy\,beam$^{-1}$. We also imaged the A-configuration data using robust +0.3 weighting (Briggs \citeyear{bri95}), which resulted in a beam of 1.05\,$\times$\,0.97 (PA\,=\,47.7$^{\circ}$) arcsec and rms noise level of 0.018 mJy\,beam$^{-1}$. The B-configuration data were imaged using robust +0.3 weighting (Briggs 1995), which resulted in a beam of 3.52\,$\times$\,3.25 arcsec (PA\,=\,53.8$^{\circ}$) and rms noise level of 0.013 mJy\,beam$^{-1}$. 

The B-configuration data revealed the radio continuum from star formation in two galaxies, the companion galaxy $\sim$11 arcsec or $\sim$20 kpc south of the ``Beetle''  and a galaxy $\sim$42 arcsec or $\sim$78 kpc NE. This allowed us to overlay the radio and optical images to within an estimated accuracy of $\sim$0.6 arcsec.

 \section{Analysis and results}
 \label{Sec:results}

 \subsection{Radio vs. optical morphology}
\label{Sec:radvsopt}

\begin{figure}
\includegraphics{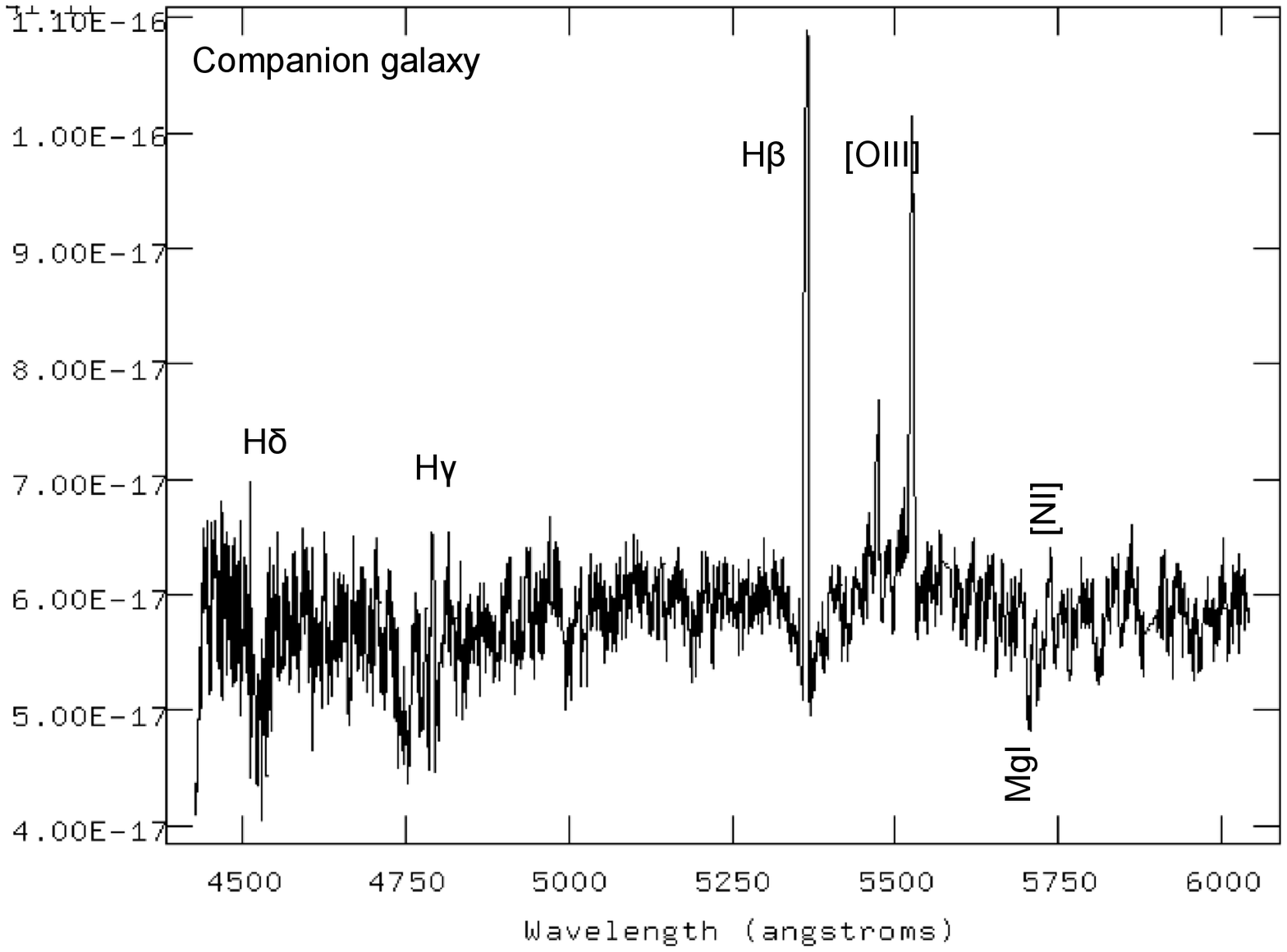}
\vspace{2.4in}
\includegraphics{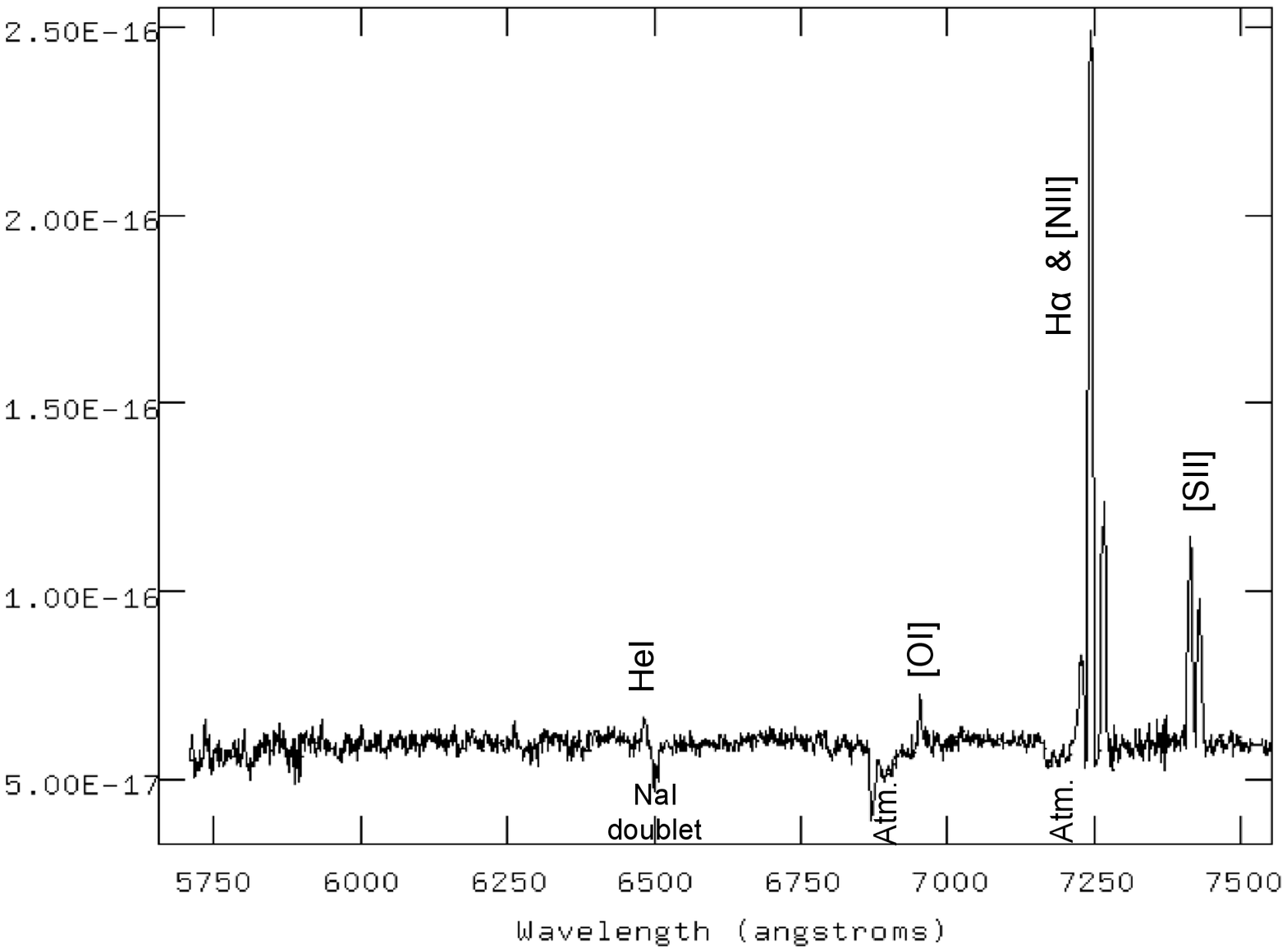}
\vspace{2.4in}
\caption{Spectra (observed frame) of the star forming companion galaxy. Prominent emission and absorption features are identified. ``Atm." marks atmospheric absorption bands. The flux is in erg s$^{-1}$ cm$^{-2}$ \AA$^{-1}$.}             
\label{fig:companion}
\end{figure}

The continuum image of the ``Beetle'' (Fig. \ref{fig:gtc}, left) reveals an elliptical galaxy with clear signatures of a merger/interaction event, such as shells and filaments. The shells do not completely encircle the central galaxy and their main axis is aligned with the galaxy major axis. This has been observed in other  elliptical galaxies  with high ellipticity (e.g. Athanassoula \&  Bosma \citeyear{ath85}, Duc et al. \citeyear{duc15}) as is the case for the ``Beetle''  ($e=b/a$=0.78, Vizier catalogue based on Hakobian et al. \citeyear{hak09}). Simulations of galactic interactions show that such shells may result from (nearly) head-on collisions with smaller disk galaxies (e.g. Quinn \citeyear{quinn84}, Struck et al. \citeyear{str99}).   The interacting galaxy is probably  the smaller    galaxy   located $\sim$11 arcsec or $\sim$20   kpc South of the QSO2\footnote{All quoted distances and sizes will be projected values unless otherwise specfied} (Fig. \ref{fig:gtc}).
The GTC-OSIRIS optical spectra of this object show that it is at the same $z$=0.10351$\pm$0.00008 as  the ``Beetle''  (Fig. \ref{fig:companion}). Based on the [OIII]/H$\beta$=0.70$\pm$0.05, [NII]/H$\alpha$=0.34$\pm$0.03, [SII]/H$\alpha$=0.46$\pm$0.03 and [OI]/H$\alpha$=0.06$\pm$0.01 ratios, it is classified as a star forming galaxy (Baldwin et al. \citeyear{bal81}, Kewley et al. \citeyear{kew06}).

\begin{figure*}
\includegraphics{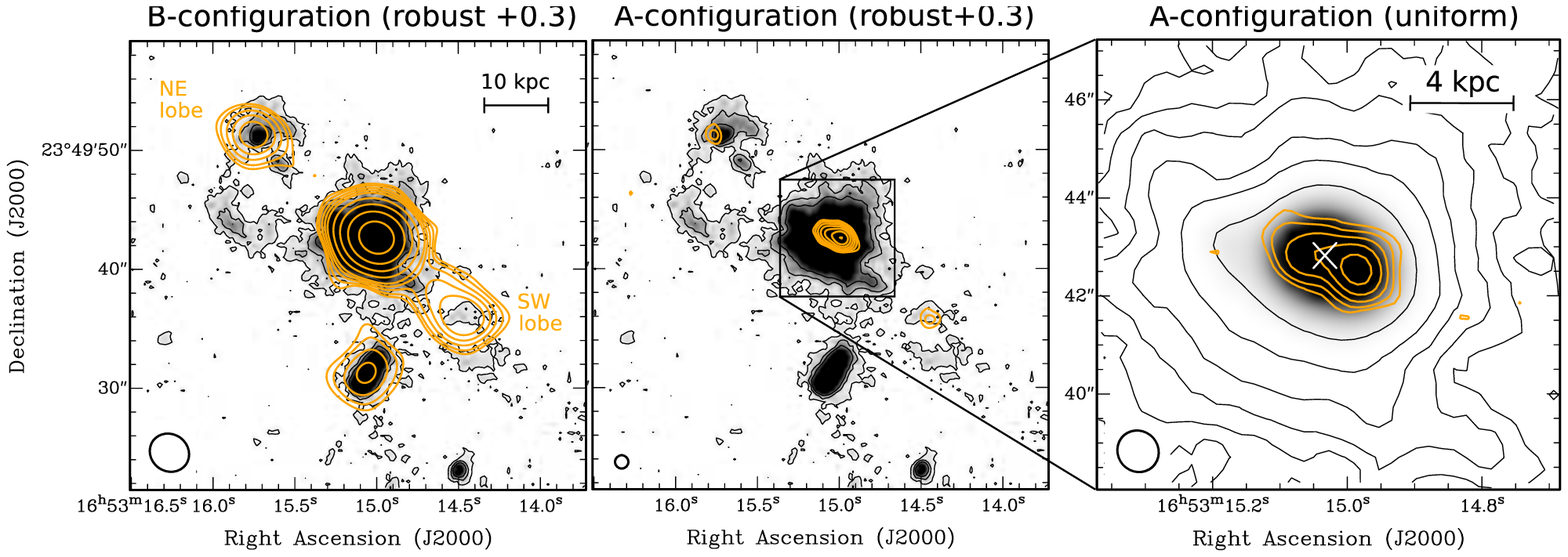}
\vspace{2.8in}
\caption{GTC H$\alpha$ (continuum subtracted) image (grey) and VLA 1.4 GHz continuum contours (orange) of the ``Beetle'' . Left: VLA B-configuration data with robust +0.3 weighting. Contour levels: 0.063, 0.085 0.13 0.18, 0.25, 0.50, 1.0, 2.0, 4.0 mJy beam$^{-1}$. Middle: VLA A-configuration data with robust +0.3 weighting. Contour levels: 0.08, 0.16, 0.32, 0.64, 1.3, 2.5, 5.0 mJy beam$^{-1}$. The outer radio hot spots overlap with the head of the H$\alpha$ arcs, suggesting  that the interaction between the radio structures and the ambient gas is responsible for the morphology of the large scale circumgalactic ionized gas. Right: zoom-in of the core-region, showing the VLA A-configuration data with uniform weighting. Contour levels: 0.2, 0.4, 0.8, 1.6, 3.2 mJy beam$^{-1}$. The beam of the various VLA data sets is shown in the bottom-left of each plot.}             
\label{fig:overlay}
\end{figure*}

The H$\alpha$+continuum narrow band image    reveals  a spectacular structure of knots and arcs  (Fig. \ref{fig:gtc}, right),  with a main axis that runs perpendicular to the main galaxy axis and the complex stellar tidal debris   seen  in the continuum image. Three arcs are identified, which are not detected in the continuum image. This already suggests that they are emission line dominated features, as confirmed by the continuum subtracted H$\alpha$ image (Fig. \ref{fig:overlay}). 
One arc is located at $\sim$13.5 arcsec (25 kpc) to the NE from the QSO2 centroid. Two more are at $\sim$10.2 arcsec (19 kpc)  and $\sim$13.7 arcsec (26 kpc) SW  of the QSO2. Thus, these features are   in the circumgalactic environment.

The morphology of the arcs is reminiscent of optical bow shocks seen in some powerful radio galaxies, where the extended radio structures are interacting with the environment (e.g. Coma A, Tadhunter et al. \citeyear{tadh00}; see also Harrison et al. \citeyear{har15} for a related example of a radio quiet QSO2: the ``Teacup««). The ``Beetle''  is certainly not associated with such a luminous radio source as Coma A
(it is $\sim$300 times fainter) and is classified as radio quiet. However, based on this resemblance a scenario of  radio jets/lobes being responsible for the observed optical ionized gas features appears plausible.
This prompted us to obtain a deeper radio map   which has confirmed the existence of a previously unknown  extended radio source. 

The VLA observations reveal a radio source with a total extent of $\sim$24.5 arcsec or $\sim$46 kpc (Fig. \ref{fig:overlay}). We detected two outer hot-spots, at $\sim$14 arcsec NE and $\sim$10.5 arcsec SW from the radio core, respectively. Therefore  the two hot spots  overlap with the NE and inner SW arcs.

Fig. \ref{fig:overlay} (middle panel)  shows our VLA high resolution (A-configuration) radio map of the  ÒBeetleÓ. The inner radio jet is detected across a total extent of  $\sim$2.3 arcsec or  $\sim$4.3 kpc (right panel). It consists of a bright hot spot $\sim$0.8 arcsec or $\sim$1.5 kpc towards the SW and fainter emission towards the NE. If this is a result of Doppler-boosting, it suggests that the inner jet is approaching on the SW side and receding on the NE side. This is consistent with the outer lobe shown in the low-resolution map (B-configuration,  Fig. \ref{fig:overlay}, left).  Only on the SW side, we detect an outer radio lobe that connects the inner radio emission to the outer SW hot-spot. This indicates that the SW side of the radio source is Doppler boosted and approaching also on large scales. Interestingly, while the modest radio power of the ``Beetle" is typical of  FR-I sources, the hot-spot morphology  resembles more closely that of the Fanaroff-Riley type II (FR-II) sources  (Fanaroff \& Riley \citeyear{fan74}).

 Our current data do not reveal whether the radio core represents re-started activity, or forms a continuous structure with the outer radio lobes and hot-spots. However, the inner  jet has  PA 68$\degr$. If we extrapolate its axis towards the NE,  it seems to cross the northern H$\alpha$ arc  at a location where this appears broken. This may suggest that the radio core is actively feeding the outer lobes. The angle between the inner jet and the outer radio axis is 19$\degr$, which would indicate precession of the radio jet or bending as it propagates outward. A deeper radio map might reveal an fainter, so far unseen extended radio structure that  may have perforated the northern arc.

The radio and emission line morphologies are  closely correlated. The large-scale radio axis is the same as the axis defined by the  brightest H$\alpha$ knots. In fact, the NE radio hot spot overlaps with the brightest circumgalactic H$\alpha$ feature.  The centre of the southern arc closer to the galaxy overlaps with the SW radio hotspot.   The close correlation between the radio and H$\alpha$ morphologies suggests that the interaction of the radio structures with the circumgalactic medium shapes the  morphology of the  large scale ionized gas, possibly also of the radio source (alternative, less convincing scenarios will be discussed in Sect. \ref{Sec:discussion}).

 While the NE  radio hotspot 
seems to define roughly  the edge of the NE ionized structures, 
the optical line emission extends well beyond  the SW radio hot spot. This will be further discussed below based on the spectroscopic data.

\subsection{The nuclear outflow}
\label{Sec:nuclear}

\subsubsection{Physical properties and excitation mechanism}
\label{Sec:nucphys}

\begin{figure*}
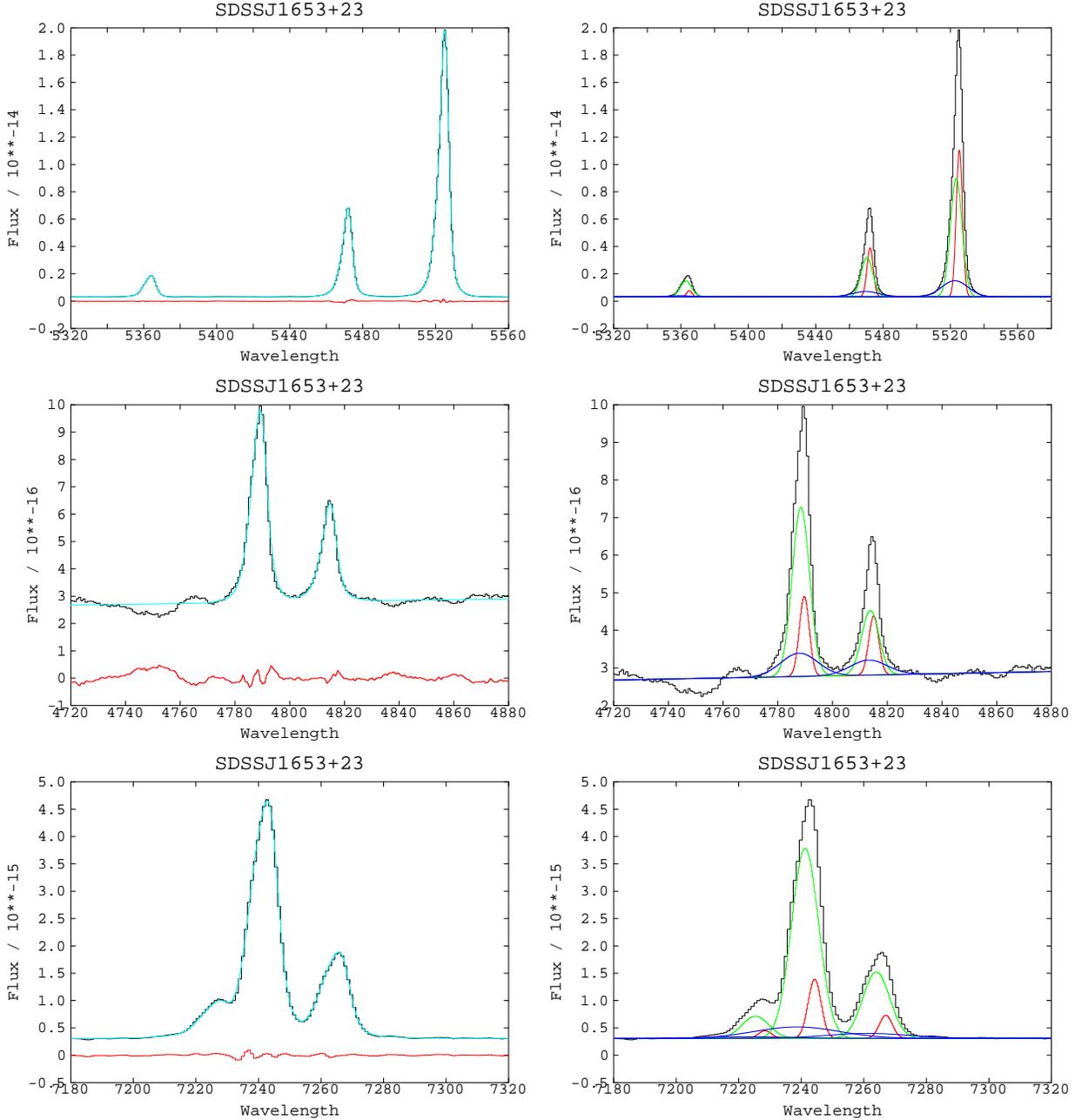

\includegraphics{fitoiiihb.ps}
\includegraphics{fitoiiihb-ind.ps}
\vspace{2.2in}
\includegraphics{fithgoiiit.ps}
\includegraphics{fithgoiiit-ind.ps}
\vspace{2.2in}
\includegraphics{fitniiha.ps}
\includegraphics{fitniiha-ind.ps}
\vspace{2.25in}
\caption{Fits of the main nuclear emission lines.  From top to bottom: [OIII] and H$\beta$, H$\gamma$ 
and [OIII]$\lambda$4363, H$\alpha$ and [NII]$\lambda\lambda$6548,6583.
The data, the fits and residuals are shown in black, cyan
and red respectively (left panels).   The individual components are shown in the right panels with different colours. Red (narrow component),  green (intermediate) 
 and blue  (broad) are used  from the most redshifted to the more blueshifted components.}
\label{fig:figfitnuc}
\end{figure*}

The results of the fits of the nuclear emission lines are shown in  Table \ref{tab:fitsnuc} and Fig. \ref{fig:figfitnuc}. Three kinematic components are isolated in the  [OIII]$\lambda\lambda$4959,5007 lines with FWHM=144$\pm$3, 398$\pm$36 and 973$\pm$25 km s$^{-1}$ respectively. The broadest component traces the ionized outflow. It is blueshifted by -144$\pm$6 km s$^{-1}$,  relative to the narrowest component. These values are in  agreement with those obtained by  \cite{vm14} based on the SDSS optical spectrum. The fits to the rest of the main optical lines produce  consistent results within the errors. 
 As already pointed out by VM14, the line ratios place the three kinematic components, including the outflow, in the AGN area of the diagnostic diagrams (Baldwin et al. \citeyear{bal81}, Kewley et al. \citeyear{kew06}). 
 
 \begin{figure*}
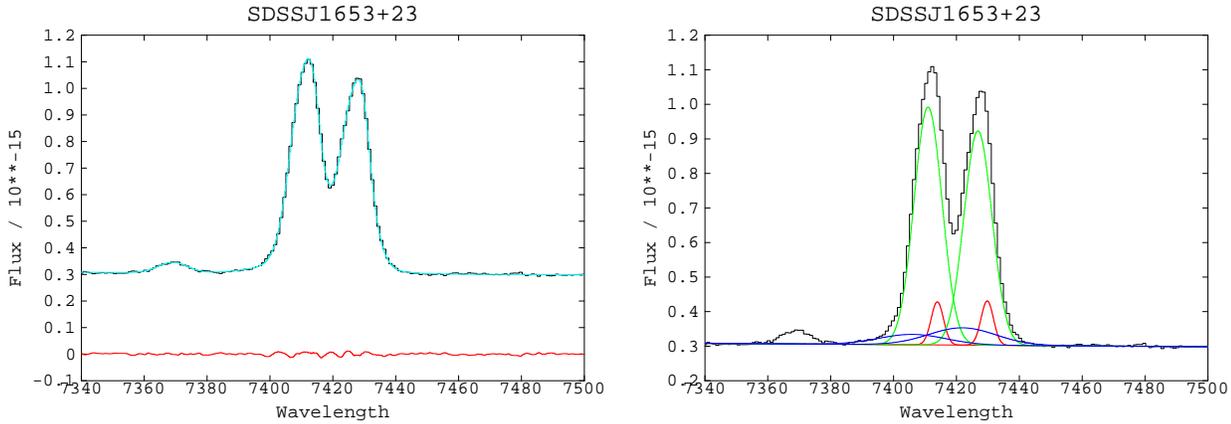

\includegraphics{fitsii.ps}
\includegraphics{fitsii-ind.ps}
\vspace{2.2in}
\caption{Fit of the [SII]$\lambda\lambda$6716,6731 doublet.  The broadest component has high density with [SII]$\lambda$6716/$\lambda$6731=0.69$\pm$0.20. The error includes the uncertainty due to  the range of acceptable fits. Colour code as in Fig.  5.}
\label{fig:figfitsii}
\end{figure*}

    \begin{table*}
\centering
\caption{Results of the  nuclear fits. The properties of the individual kinematic components are
shown. The velocity shifts $V_{\rm s}$ have
been calculated relative to the $\lambda$ of the narrowest component. Numbers in italics correspond to lines for which the kinematic parameters have been fully constrained with other emission lines (hence no errors).} 
\begin{tabular}{llll}
\hline
Narrow component &  FWHM & $V_{\rm s}$  & Flux \\
	&  km s$^{-1}$  &  km s$^{-1}$    & $\times$10$^{-15}$ \\
	& 	&  	& erg s$^{-1}$   \\   
\hline
~[OIII]$\lambda$5007 & 144$\pm$3	& 0$\pm$3	& 49.8$\pm$0.7   \\ 
~H$\beta$ & 127$\pm$25&  0$\pm$3	&  2.2$\pm$0.4	 \\ 
~H$\gamma$ &	{\it 144} & 0$\pm$3	& 1.2$\pm$0.1	 \\ 
~[OIII]$\lambda$4363  & {\it 144} & 0$\pm$3	& 0.8$\pm$0.1		 \\  \hline
~H$\alpha$ &  113$\pm$11 & 0$\pm$3	 & 5.6$\pm$0.3 \\ 
~[NII]$\lambda$6583 & {\it 113$\pm$11} & {\it 0$\pm$3} &  2.3$\pm$0.1\\
~[SII]$\lambda$6716 & 91$\pm$18 & 0$\pm$4 & 0.6$\pm$0.1 \\  
~[SII]$\lambda$6731  & {\it 91} & {\it 0}  & 0.6$\pm$0.1\\  \hline
Intermediate component &  FWHM & $V_{\rm s}$  & Flux \\
\hline
~[OIII]$\lambda$5007 & 398$\pm$36 	& -105$\pm$3	& 74.5$\pm$0.3   \\ 
~H$\beta$ & 398$\pm$13 	&  -105$\pm$5	&  9.2$\pm$0.3 \\ 
~H$\gamma$ &	{\it 398} 	& -91$\pm$10	& 3.2$\pm$0.2	 \\
 ~[OIII]$\lambda$4363  & {\it 398} 	& -91$\pm$10	 & 1.3$\pm$0.2		 \\  \hline
~H$\alpha$ &  387$\pm$3 & -126$\pm$3	 & 37.5$\pm$0.5 \\ 
~[NII]$\lambda$6583 & {\it 387$\pm$3} & {\it -126$\pm$3} &  13.0$\pm$0.3\\
~[SII]$\lambda$6716 & 392$\pm$4 & -119$\pm$3 & 7.7$\pm$0.2 \\  
~[SII]$\lambda$6731  & {\it 392} & {\it -119} & 6.9$\pm$0.2\\  \hline
Broad component &  FWHM & $V_{\rm s}$  & Flux \\
\hline
~[OIII]$\lambda$5007 & 973$\pm$25 	& -144$\pm$6	& 22.9$\pm$0.9   \\ 
~H$\beta$ & 942$\pm$78 	&  -195$\pm$84&  1.3$\pm$0.2	 \\
 ~H$\gamma$ &	{\it 973} & -128$\pm$16	& 1.1$\pm$0.3	 \\
 ~[OIII]$\lambda$4363  & {\it 973} & -128$\pm$16 & 0.7$\pm$0.2		 \\  \hline
~H$\alpha$ &  1267$\pm$174 & -235$\pm$40	 & 7.4$\pm$0.8 \\ 
~[NII]$\lambda$6583 & {\it 1267$\pm$174} & {\it -235$\pm$40} &  3.0$\pm$0.7\\
~[SII]$\lambda$6716& 1000$\pm$75 &  -320$\pm$56 & 0.9$\pm$0.2 \\  
~[SII]$\lambda$6731 & {\it 1000} & {\it -320} & 1.3$\pm$0.3 \\  \hline
\end{tabular}
\label{tab:fitsnuc}
\end{table*}

\begin{table*}
\centering
\caption{Line ratios (not corrected for reddening) and physical properties of the nuclear total spectrum and the individual kinematic components. The electron temperature $T_e$ has been calculated using the reddening corrected [OIII]4959,5007/[OIII]4363 ratio.} 
\begin{tabular}{cllll}
\hline
Ratio &  Narrow  &  Intermediate  & Broad & Total \\ \hline
~[OIII]5007/H$\beta$ & 22.2$\pm$4.3 & 8.1$\pm$0.3  &  17.8$\pm$3.3  & 11.9$\pm$0.3\\
~H$\gamma$/H$\beta$ & 0.54$\pm$0.12 & 0.35$\pm$0.03  & 0.84$\pm$0.30 & 0.43$\pm$0.08 \\
~[OIII]4959,5007/[OIII]4363 & 84$\pm$11 & 76$\pm$12 & 54$\pm$11  & 73$\pm$7 \\  
~H$\alpha$/H$\beta$ &  2.6$\pm$0.5 & 4.1$\pm$0.2 & 5.8$\pm$1.2 & 3.98$\pm$0.07\\
~[NII]6583/H$\alpha$ & 0.41$\pm$0.04 & 0.35$\pm$0.01 & 0.40$\pm$0.10 & 0.36$\pm$0.02\\
~[SII]6716,6731/H$\alpha$ & 0.22$\pm$0.02 & 0.39$\pm$0.01& 0.29$\pm$0.05 & 0.36$\pm$0.01 \\
~[SII]6716/[SII]6731& 0.99$\pm$0.15 & 1.11$\pm$0.04 & 0.69$\pm$0.20 & 1.04$\pm$0.05  \\  \hline  
~$n_e$ (cm$^{-3}$) & 730$^{+634}_{-365}$ & 425$^{+89}_{-78}$ & 2571$^{+14181}_{-1461}$ & 588$^{+111}_{-223}$ \\ 
~E(B-V) & 0 & 0.33$\pm$0.04 & 0.63$\pm$0.18  & 0.30$\pm$0.02  \\
~$T_e$ ($\times$10$^4$ K) & 1.38$^{+0.08}_{-0.07}$    & 1.56$^{+0.19}_{-0.13}$   & 2.2$^{+4.6}_{-0.5}$ & 1.62$^{+0.10}_{-0.08}$ \\
\hline
\end{tabular}
\label{tab:fitsnuc2}
\end{table*}

 \begin{figure*}
\includegraphics{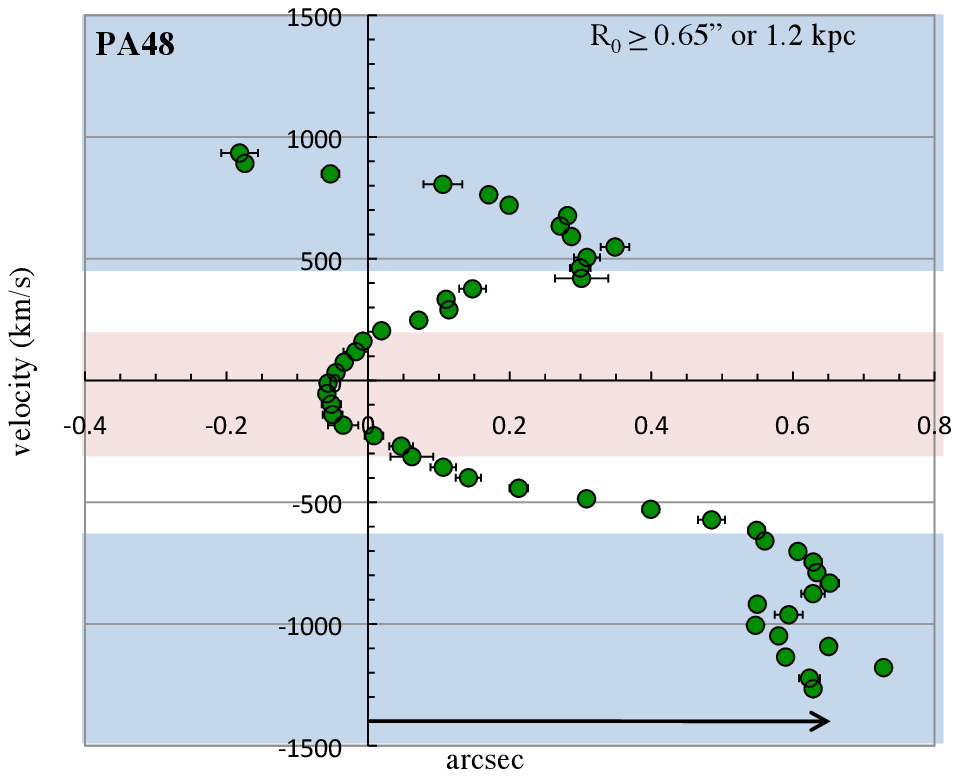}
\includegraphics{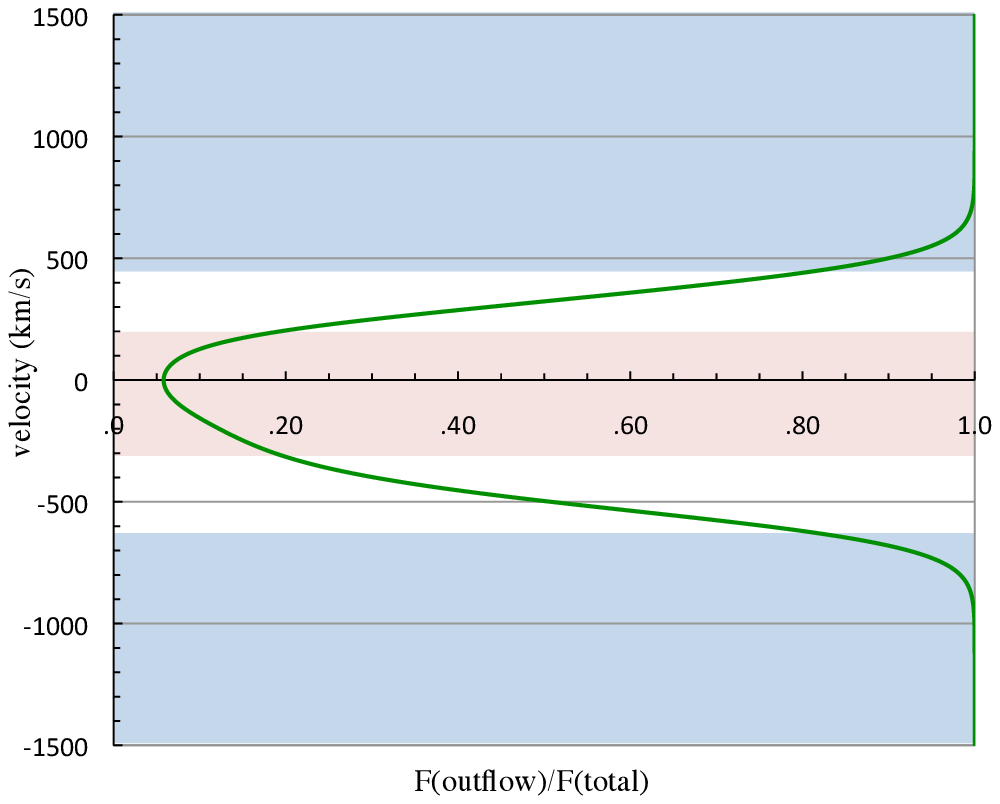}
\vspace{2.85in}
\includegraphics{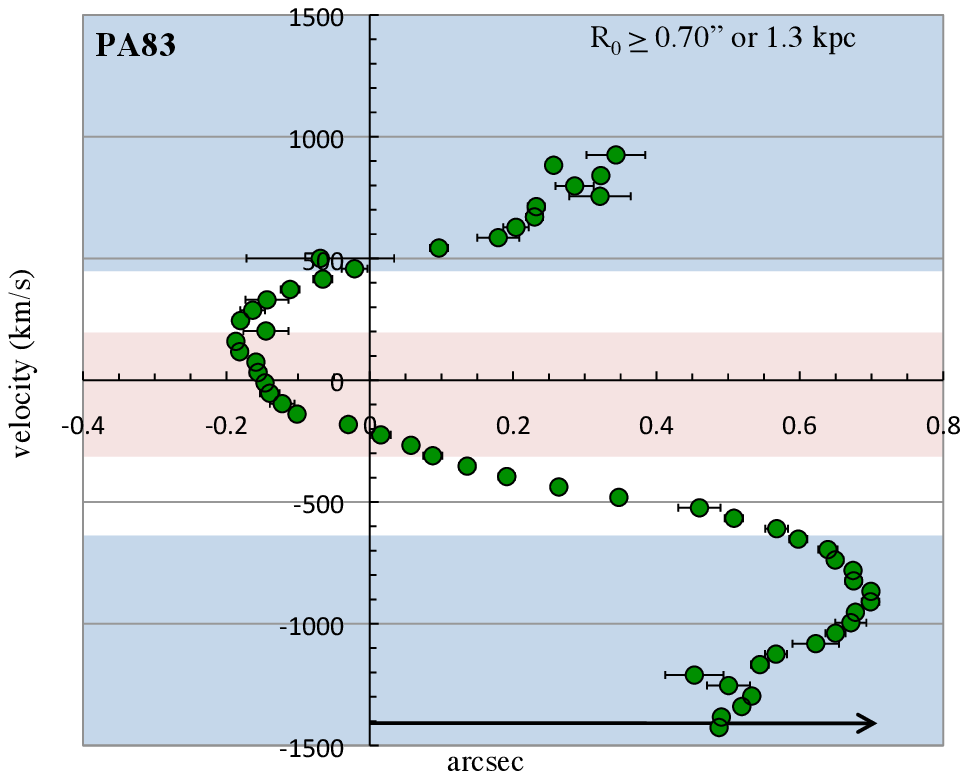}
\includegraphics{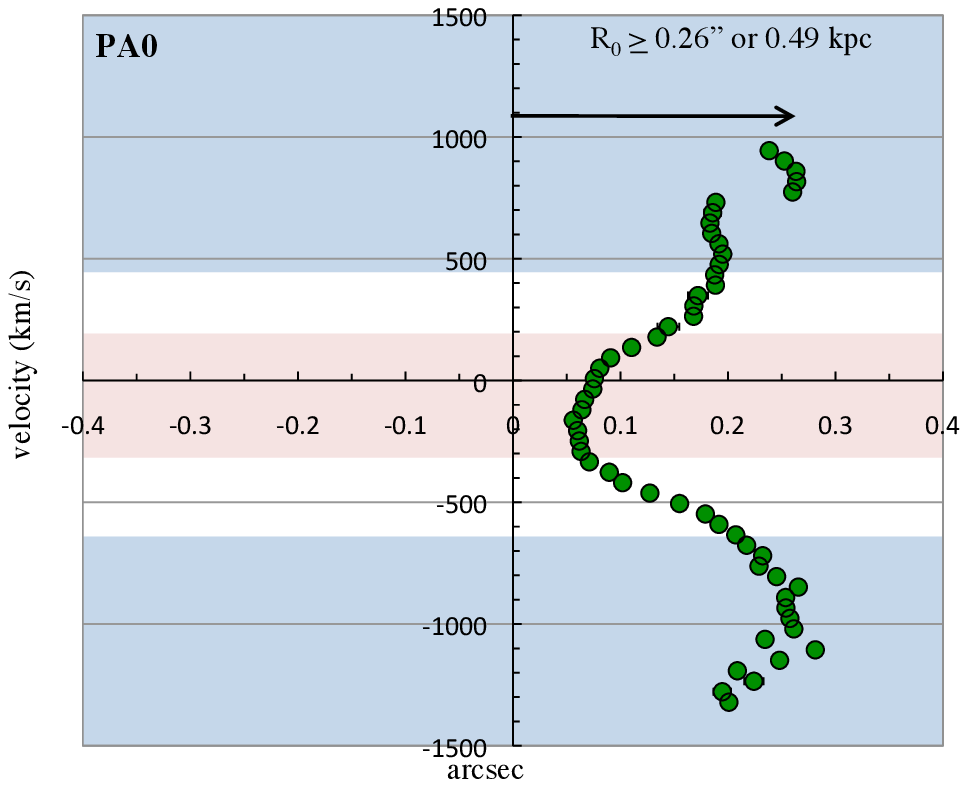}
\vspace{2.85in}
\caption{Spectroastrometric analysis   following Villar Mart\'\i n et al. (2016)  along the radio hot spots axis PA48,   PA83  and   PA0. Left panels and bottom right: Shift of the spatial centroid measured for [OIII] 
at different velocities relative to the continuum centroid along the three PA. Top-right: Velocity shift versus the relative contribution of the outflowing gas to the total [OIII] line flux $\frac{F_{\rm outflow}}{F_{\rm total}}$.  The blue areas mark the range of velocities for which the outflow dominates the line flux ($\frac{F_{\rm outflow}}{F_{\rm total}}\ge$0.8). The red areas mark the velocities for which its contribution is minimum ($\frac{F_{\rm outflow}}{F_{\rm total}}\le$0.2). The emission at  velocities dominated by the outflow are
in general shifted relative to the continuum centroid. The arrow marks the assumed spatial shift representing the lower limit for the radial size of the outflow $R_{\rm o}$. The top-right plot  is almost identical along the three slit PAs.}
\label{astrom1653}
\end{figure*}

The reddening E(B-V), electron density $n_{\rm e}$ and temperature $T_{\rm e}$ have been inferred  for the three kinematic components using H$\alpha$/H$\beta$ (H$\gamma$/H$\beta$ is affected by large uncertainties especially for the outflowing gas; see Fig. \ref{fig:figfitnuc}),  [SII]$\lambda$6716/$\lambda$6731 and reddening corrected [OIII]$\lambda\lambda$4959,5007/$\lambda$4363 respectively (Table \ref{tab:fitsnuc2}).

Inferring $n_{\rm e}$ for the outflow is not trivial, given the complexity of the [SII] doublet (3 components per line) and the partial blend of both lines (Fig.  \ref{fig:figfitsii}). It is found that the fits producing kinematically reasonable results (in comparison with the other emission lines) and physically meaningful results (as opposite to unphysical, such as [SII]$\lambda$6716/$\lambda$6731$>$1.4 for any of the individual components) result in    [SII]$\lambda$6716/$\lambda$6731=0.69$\pm$0.20, as indicated in  Table \ref{tab:fitsnuc2}.  This implies $n_{\rm e}=$2571$^{+14181}_{-1461}$  cm$^{-3}$ for the outflowing gas. The minimum possible value of the [SII] ratio (0.49)  is actually in the regime where the ratio saturates and becomes insensitive to increasing densities (Osterbrock \citeyear{ost89}). Thus,  the ``best'' density $n_{\rm e}=$2571 cm$^{-3}$ is effectively almost a lower limit.
The outflowing gas density is  substantially higher than the densities of the narrow ($n_{\rm e}=$730$^{+634}_{-365}$  cm$^{-3}$)  and the intermediate  ($n_{\rm e}=$425$^{+89}_{-78}$  cm$^{-3}$) components respectively.  
  Different works suggest that such high densities are frequent in the nuclear ionized outflows of luminous type 2 AGN (e.g. Holt et al. \citeyear{holt11}, \cite{vm14}, Villar Mart\'\i n et al. \citeyear{vm15}).

We have tested whether a good fit can also be obtained by forcing the broadest component to have  low density ([SII]$\lambda$6716/$\lambda$6731=1.4 or $n_{\rm e}\la$100 cm$^{-3}$, Osterbrock \citeyear{ost89}). However, we get reasonable fits  only  under very strict constraints on the kinematics and individual line ratios. We thus consider that the higher densities are more realistic. 
This is supported also by the high [OIII]/H$\beta$=17.8$\pm$3.3 ratio, which points to $n_{\rm e}\sim$few$\times$1000-10$^{4}$ cm$^{-3}$ (\cite{vm14}).  High densities in the narrow line region (NLR) of the ``Beetle'', where 
the gas emitting the broadest component is likely to be located, are also suggested by [ArIV]$\lambda$4711/$\lambda$4740=1.10$\pm$0.08 which implies $n_{\rm e}$=3450$^{+1280}_{-1127}$, i.e. at least $n_{\rm e}\sim$2300 cm$^{-3}$.

The nuclear outflow   shows  the highest reddening  of the three kinematic components with H$\alpha$/H$\beta$=5.8$\pm$1.2, as often found for the ionized outflows in QSO2 (\cite{vm14}). As proposed by \cite{vm15}, the trend of the outflowing gas to have the highest reddening and density  can be naturally explained if the outflow is concentrated in a smaller region, closer to the the AGN than the intermediate and narrow components  (Bennert et al. \citeyear{ben06a},\citeyear{ben06b}). 
It also shows significantly higher $T_ {\rm e4} = \frac{T_ {\rm e}}{\rm 10^4}$=2.2$^{+4.6}_{-0.5}$ K, compared with 1.38$^{+0.08}_{-0.07}$    and 1.56$^{+0.19}_{-0.13}$  K  for the narrow and intermediate components respectively. The additional heating may be produced by shocks induced by the outflow (e.g. Villar Mart\'\i n et al. \citeyear{vm99}).

\subsubsection{Radial size}
\label{Sec:size}

We have derived the radial size  $R_ {\rm o}$ of the nuclear outflow along the three different slit PA. Upper limits or actual sizes are constrained by comparing the spatial distribution of the broadest kinematic component  and the seeing profile.
 Lower limits are obtained by applying the spectroastrometric method (see  Carniani et al. \citeyear{car15} and Villar Mart\'\i n et al. \citeyear{vm16} for a detailed description and discussion on the uncertainties of both methods).  
 
 The  method of spectroastrometry, which gives lower limits, consists of measuring the relative position of the  centroid of the [OIII] spatial profile as a function of velocity.  We assume as spatial zero the continuum centroid. Although this does not necessarily mark the exact location of the AGN,  it is a reasonable assumption for our purposes, since the associated uncertainty on $R_{\rm o}$ is expected to be $\la$few$\times$100 pc. We consider as zero velocity that of the narrow core of the [OIII] nuclear line.  The results are shown in Fig. \ref{astrom1653}. We infer $R_ {\rm o}\ga$1.2 kpc, $\ga$1.3 kpc and $\ga$0.5 kpc along PA48, PA83 and
PA0 respectively.

 \begin{figure}
\includegraphics{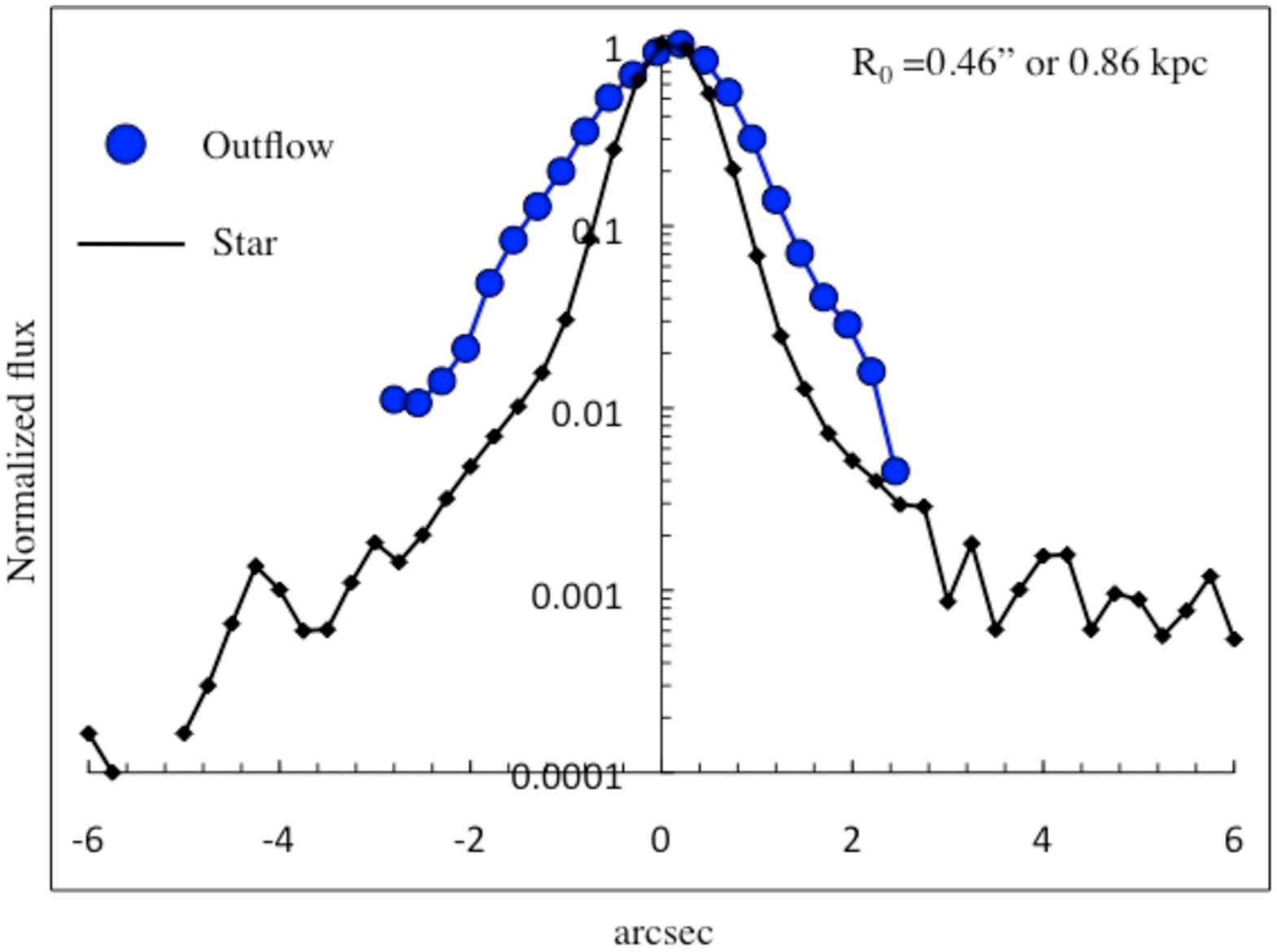}
\vspace{2.5in}
\caption{Comparison of the nuclear outflow spatial distribution along PA48 with the seeing profile. The outflow is spatially resolved and has $R_ {\rm o}$=0.86$\pm$0.07 kpc, corrected for seeing broadening.} 
\label{spatbroadob1}
\end{figure}

To compare the  seeing profile with the spatial distribution of the outflow, a spatial profile of the [OIII] line (continuum subtracted) was extracted from  a spectral (velocity) window clearly dominated by the outflow. As discussed in \cite{vm16}, it is essential to use an accurate seeing profile for a reliable assessment of any possible excess above the seeing distribution, especially the wings. Following that work, the seeing profiles along PA48, PA83 and PA0 were specifically built  using non-saturated stars with well detected wings in the acquisition images obtained at similar time as the spectra. The seeing profiles were extracted from apertures of the same width in arcsec as the narrow slit of the spectroscopic observations. 

The outflow spatial distribution is found to be resolved  (FWHM$_ {\rm obs}$=1.25\arcsec$\pm$0.05\arcsec) relative to  the seeing  (FWHM=0.85\arcsec$\pm$0.04\arcsec) only along PA48 (Fig. \ref{spatbroadob1}).  We obtain $R_ {\rm o}$=0.86$\pm$0.07 kpc after correcting for seeing broadening. The error accounts for the uncertainty on FWHM$_{\rm obs}$ and the seeing FWHM. All uncertainties considered, this $R_ {\rm o}$ is in good agreement with the spectroastrometric result ($R_ {\rm o}\ga$1.2 kpc). 
\begin{figure*}
\includegraphics{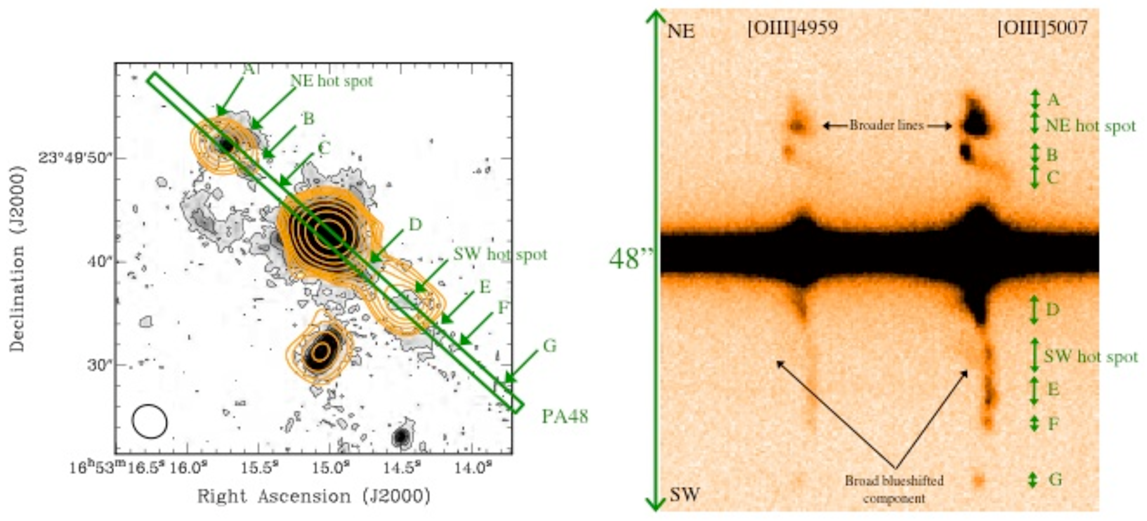}
\vspace{3.8in}
\caption{Left: H$\alpha$ (continuum subtracted) image  and  VLA radio map overlay (see Fig. 4). The  PA48 slit is indicated and
 the main emission line features isolated in the
long slit spectrum. Right: GTC long-slit spectrum covering the [OIII]$\lambda\lambda$4959,5007 spectral window. 
1-dim spectra were extracted from the apertures  indicated in the figure. The lines show distinct kinematics at the location of the radio hot spots. They are broader coinciding with  the NE hot spot and a broad blueshifted  wing is detected at the location of the SW hot spot.}             
\label{ap48}
\end{figure*}

\begin{figure*}
\includegraphics{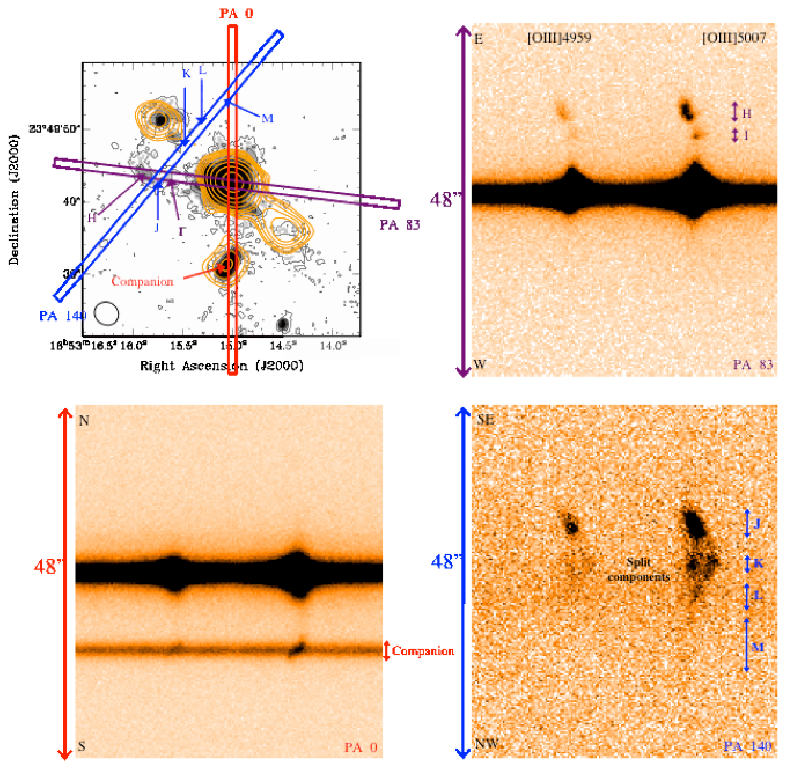}
\vspace{6.5in}
\caption{Top-left:  The slits PA83, PA0 and PA140 (avoiding the nucleus) are   shown.  The  apertures used in our study are indicated in the 2-dimensional [OIII] spectra. Notice the split components in aperture K along PA140, at the location 
of the radio axis (bottom right panel).}               
\label{apotherpa}
\end{figure*}

\begin{figure*}
\includegraphics{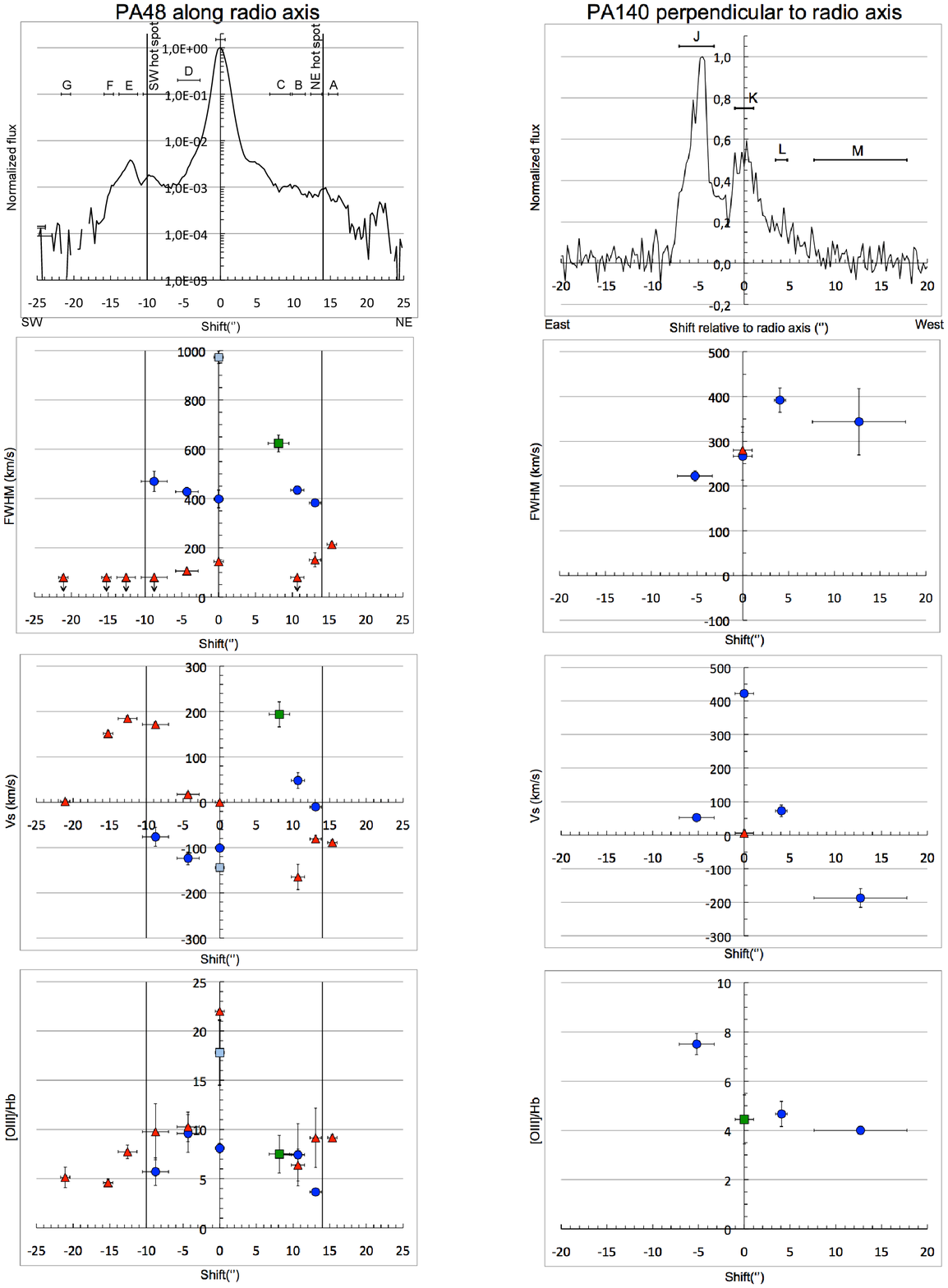}
\vspace{8in}
\caption{Spatial variation of the kinematic and ionization properties of the ionized gas. Left: From top to bottom the panels show the spatial distribution of the [OIII] flux, FWHM, $V_{\rm s}$ and [OIII]/H$\beta$ along the radio axis (left). Blue is used for the broadest component (outflowing gas) and red for the narrowest (ambient gas). Right: Same plots, but along PA140. The spatial zero marks the location of the radio axis. Green squares are used when the lines ([OIII] and/or H$\beta$) appear double peaked but only one component could be fitted due to the low S/N of the spectrum. The capital letters on the top panel represent the apertures identified in Figs. 9 and 10.}
\label{fig:kinem}
\end{figure*}

 The same method demonstrates that the outflow spatial distribution is consistent with the seeing disk along both PA83 (seeing FWHM=1.25\arcsec$\pm$0.07\arcsec) and  PA0 (seeing FWHM=1.1\arcsec$\pm$0.1\arcsec). This implies  $R_ {\rm o}\la$0.7 and 0.8 kpc respectively, also in reasonable  agreement with the spectroastrometric lower limits taking into account all uncertainties.

 \subsubsection{Mass and energetics}
\label{Sec:mass}

 Following standard procedures (see Villar Mart\'\i n et al. \citeyear{vm16} for a discussion on the method and all related uncertainties) we have calculated the outflow mass $M_{\rm o}$, mass outflow rate $\dot M_{\rm o}$, energy $\dot E_{\rm o}$ and momentum $\dot p_{\rm o}$ injection rates. We assume $n_e=$2571 cm$^{-3}$, $V_{\rm o} = |V_{\rm s}|$+ FWHM/2=630 km s$^{-1}$ (see Tables \ref{tab:fitsnuc} and  \ref{tab:fitsnuc2}; Arribas et al. \citeyear{arr14}) and 
 $R_ {\rm o}$=0.86 kpc. The H$\beta$ flux of the outflowing component isolated in the nuclear spectrum is F(H$\beta_{\rm o}$)=(1.3$\pm$0.2)$\times$10$^{-15}$ erg s$^{-1}$. Slit losses   are expected to be small (F(H$\beta_{\rm o}$)=(1.5$\pm$0.2)$\times$10$^{-15}$ erg s$^{-1}$ as measured from the SDSS 3" fibre spectrum, \cite{vm14}). Correcting  for reddening (as implied by H$\alpha$/H$\beta\sim$5.8), the outflow luminosity is
  L(H$\beta_{\rm o}$)$\sim$3.2$\times$10$^{41}$ erg s$^{-1}$.  Therefore, $M_{\rm o}$=3.5$\times$10$^{6}$ M$_{\odot}$ and $\dot M _{\rm o}$=2.6 M$_ {\odot}$ yr$^{-1}$. In addition,  $\dot E_{\rm o}$=3.2$\times$10$^{41}$ erg s$^{-1}$ (or $\sim$10$^{-5} \times L_{\rm Edd}$) and $\dot p_{\rm o}$=1.0$\times$10$^{34}$ dyne (or $\sim$0.01 $L_{\rm Edd}$/c), where $L_{\rm Edd}$ is the Eddington luminosity (Sect. \ref{Sec:object}).  

 As discussed in \cite{vm16}, the outflowing gas is likely to have a density gradient with increasing densities at decreasing distances from the  AGN. If such is the case, the assumption of higher densities would result in   even lower values.
 
\subsection{The circumgalactic gas}
\label{Sec:ext}

We have studied the spatial distribution and properties of the ionized gas along the four  slit PA  described in Sect. \ref{Sec:gtcspec}. 
1-dimensional spectra were extracted from different apertures centred on prominent emission line features detected along each slit and low surface brightness regions in between. The slit locations, the 2-dimensional spectra of the [OIII] doublet and the apertures used for our study are shown in Figs. \ref{ap48} and \ref{apotherpa}.

\subsubsection{Spatial extension of the emission lines}
\label{Sec:extension}

Extended ionized gas is detected along the 4 slit PA (Table \ref{tab:extensions}). The maximum extension  is measured along PA48 (i.e. the radio hot spots axis, Fig. \ref{ap48}), with line emission  detected across  $D_{\rm max}\sim$38 arcsec or 71 kpc.  Towards the NE, as already seen in the H$\alpha$ image,  the brightest [OIII] emission overlaps with the NE hot spot. Towards the SW, line emission is detected well beyond the  hotspot.  A compact (along the slit) emission line feature is detected at $\sim$10.5 arcsec  or $\sim$20 kpc beyond  the hot spot and $R_{\rm max}\sim$41 kpc from the QSO2 centroid (aperture (ap.) G in Fig.  \ref{ap48}).   Line emission is detected also between the high surface brightness features, at least from ap. A to  F. 

Along PA83 the spatial distribution of the emission lines is highly asymmetric. $D_ {\rm max}\sim R_ {\rm max}\sim$23 kpc towards the E,  up to the Eastern edge of the H$\alpha$ bubble identified in the narrow band image (ap. H, Fig.  \ref{apotherpa}). Towards the W, the lines are just barely resolved compared with the seeing disk.

Along PA0,  extended line emission associated with the QSO2 is tentatively detected only towards the N,  up to $R_{\rm max}\sim$29 kpc. This detection is confirmed by the PA140 spectrum (Fig.  \ref{apotherpa}). Along this PA, which did not cross the galactic nucleus, the line emission extends at both sides of the radio axis across $D_ {\rm max}\sim$23 arcsec or 43 kpc. 
This suggests that  line emission is detected across the entire area delineated by  the NW edge-brightened bubble.

 \begin{table}
\centering
\caption{Maximum [OIII] total spatial extension  $D_ {\rm max}$ and maximum extension $R_ {\rm max}$ from the galaxy continuum centroid
along the four slit PA. $R_ {\rm max}$ is  quoted only for the three PA that cross the galactic nucleus. The direction  of $R_ {\rm max}$ from the continuum centroid is given in the last column.} 
\begin{tabular}{llll}
\hline
PA	&	$D_ {\rm max}$ &  $R_ {\rm max}$ & Direction \\ 
	&	kpc 	& kpc	\\ \hline
48$\degr$ 	& 71	& 41	& SW \\ 
83$\degr$ 	& 23	& 23	& E \\ 
0$\degr$ 	&  29	& 29 & N	 \\ 
140$\degr$   &	43  &	 \\ 
\hline
\end{tabular}
\label{tab:extensions}
\end{table}

\subsubsection{Kinematics}
\label{Sec:kinem}

The visual inspection of the  2-dimensional [OIII]  spectra reveals variations on the kinematic properties of the gas (FWHM and $V_ {\rm s}$) correlated with the radio structures. This is particularly striking along the radio axis (PA48) and PA140. Along  PA48  (Fig. \ref{ap48}) the lines are clearly broader at the location of the NE hot spot. Moreover, a broad blueshifted  wing is  detected at the location of the SW hot spot. A sharp, continuous change in $V_{\rm s}$ of $\sim$550 km s$^{-1}$ occurs in aperture C (Fig. \ref{ap48}, right panel)
in a region of very low surface brightness. Along PA140  the lines appear clearly split in two components at the location crossing the radio  axis   (see aperture K in Fig. \ref{apotherpa}, bottom right panel).

We show in Fig. \ref{fig:kinem} (left panels) how the kinematics of the ionized gas  varies spatially along PA48.  The lines were fitted with 1 or 2 Gaussians, as required, using the 1-dimensional spectra extracted from each aperture (Fig. \ref{ap48}). FWHM and $V_ {\rm s}$ are plotted at each location for the individual kinematic components.
 The three  components identified in the nuclear spectrum (see Sect. \ref{Sec:nuclear}) are also plotted at the spatial zero. The variation of the [OIII]/H$\beta$ ratio is  shown when 
measurable.

Two kinematic components are required to fit [OIII]  across the spatial range circumscribed by the radio hot spots (a single component is required beyond).  The narrowest component has a FWHM$\la$150 km s$^{-1}$ and extends all the way along the slit, including beyond the maximum observed spatial extend of the radio emission. In contrast, a broader component of FWHM$\sim$380-470 km s$^{-1}$ is detected across the entire spatial range circumscribed by the radio hot spots.  Such broad lines reveal turbulent kinematics.   For comparison, the extended ionized gas in mergers show typical FWHM$<$250 km s$^{-1}$ even in the most dynamically disturbed systems with signs of AGN activity (Bellocchi et al. \citeyear{bel13}, Arribas et al. \citeyear{arr14}).    

The split components are confirmed and clearly seen in the 2-dimensional PA 140 extranuclear spectrum (Fig. \ref{apotherpa}) at the location where the slit intersects the radio axis (see also Fig. \ref{fig:kinem}, right panels). 
Turbulent kinematics are  detected not only along the radio axis but also at locations far from it. As an example, FWHM=392$\pm$27  and  344$\pm$74  km s$^{-1}$ for apertures L and M along PA140, which trace circumgalactic gas at $\sim$8 and 25 kpc form the radio axis in the perpendicular direction.       Higher S/N spectra may reveal that the large FWHM are due to two split kinematic components.

 The velocity curve of the turbulent gas along the radio axis (Fig. \ref{fig:kinem}, left) shows a change in $V_{\rm s}$, $|\Delta V_{\rm s}|$=119$\pm$23 km s$^{-1}$, from blueshifted in the SW to redshifted in the NE. As discussed in Sect. \ref{Sec:radvsopt}, the radio orientation places the inner jet and SW radio hot spot closer to the observer.   The kinematic behaviour of the circumgalactic turbulent gas suggests that it is expanding in an outflow, with the NE  gas  moving  away and the SW gas moving towards the observer. 
 
The narrow component presents the opposite behaviour (from more redshifted in the SW to more blueshifted in the NE).  
The scenario where this component is emitted by gas on the far side of the expanding bubbles while the broad component is emitted by the near side seems unlikely.  It is not clear why both sides would emit   lines with different FWHM and   ionization level (see below). A more natural explanation is that the narrow component is  emitted by a gas reservoir unaffected by the radio structures. This is supported by the lack of any apparent correlation in either the spatial distribution or the kinematics of this narrow component with the radio structures. In this scenario, its velocity curve suggests that it may be infalling gas.

The ionization level as traced by [OIII]/H$\beta$ presents minimum values for the turbulent gas  at the hot spots (Fig. \ref{fig:kinem}, left), where [OIII]/H$\beta\sim$3.7-5.7 compared with $\sim$9.2-9.8 for the narrow component.

Along PA83, the eastern bubble (aperture H, Fig. \ref{apotherpa}) has FWHM=197$\pm$16 km s$^{-1}$ and [OIII]/H$\beta$=6.8$\pm$0.7. The profiles are slightly asymmetric with a red excess.  The inner knot  (aperture I) has similar FWHM=219$\pm$15 km s$^{-1}$, also with a slight red excess, and [OIII]/H$\beta$=8.4$\pm$1.2. The similarity of  the FWHM and [OIII]/H$\beta$ ratio
compared with aperture A along PA48  are consistent with these regions being part of the same structure, as shown by the H$\alpha$ image.

We shall discuss in Sect. \ref{Sec:discussion} how all the above results are natural consequences of the interaction between the radio structures with the ambient circumgalactic gas, which is compressed and kinematically perturbed in the process.

\subsubsection{Physical properties and excitation mechanism}
\label{Sec:ionization}

Different line ratios measured with the 1-dimensional spectra extracted from prominent features along PA48 and PA83 
were
used to investigate the excitation mechanism and physical properties of the circumgalactic gas. In addition to  the NE hot spot   (PA48) 
and the Eastern bubble (aperture  H, PA83, Fig. \ref{apotherpa}), a spectrum was extracted along
PA48 from a large aperture covering the location of the SW radio hot spot and apertures E and F. Relevant  line ratios are shown in Table \ref{tab:ratios-circ}. 

 The  [SII] doublet implies $n_{\rm e}=$184$^{+60}_{-54}$ cm$^{-3}$ at the NE hot spot location (Table \ref{tab:ratios-circ}). Interestingly, the temperature $T_ {\rm e}$ implied by the
[OIII] lines is very high. The [OIII]4959,5007/[OIII]4363 ratio corrected for reddening with H$\alpha$/H$\beta$  implies  $T_{\rm e}$=24,000$^{+27,800}_{-5,600}$ K.  The observed, uncorrected ratio sets a lower limit $T_{\rm e}$=19,000$^{+1,500}_{-1,200}$, also rather high.

\begin{figure*}
\includegraphics{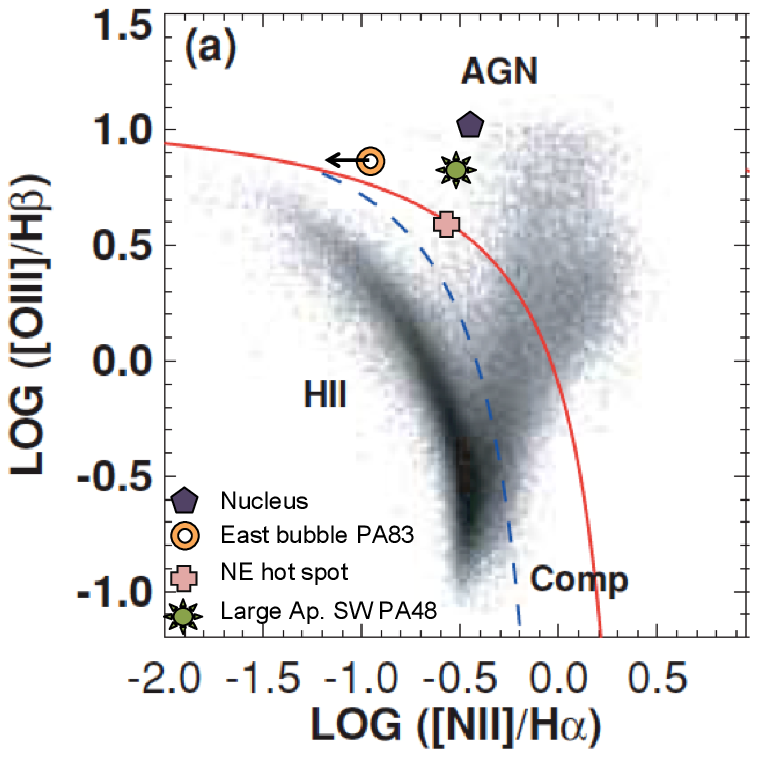}
\includegraphics{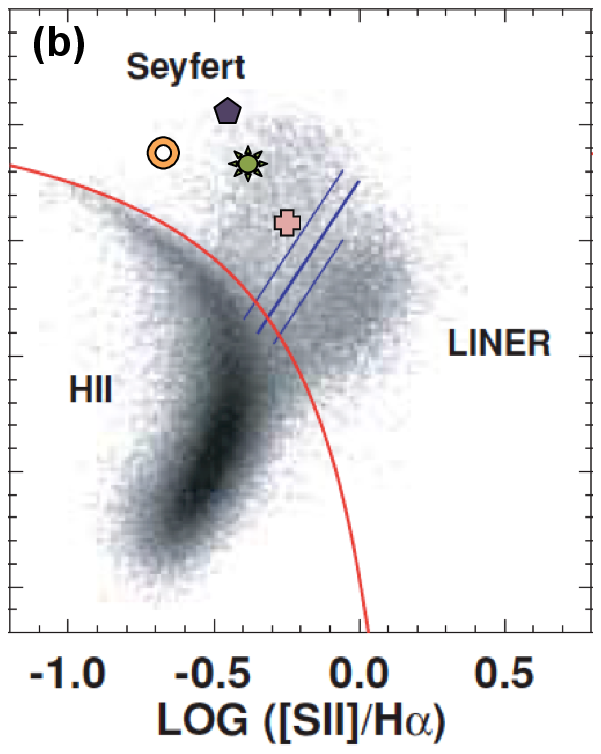}
\includegraphics{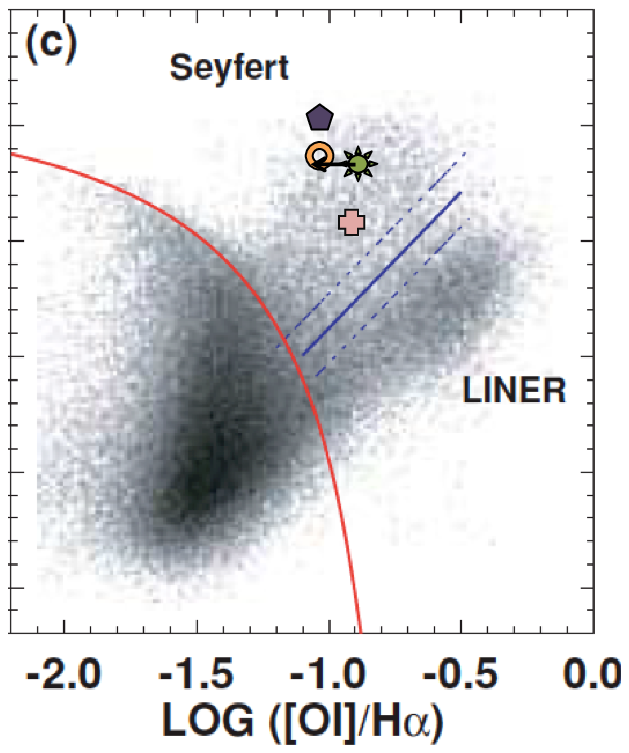}
\vspace{3.0in}
\caption{BPT  diagnostic diagrams taken from Kewley  et al. (2006) with the location of the nuclear spectrum of the ``Beetle'' ,
the NE hot spot, the eastern bubble (PA83) and the large aperture SW of the QSO2 along PA48.  Data error-bars  smaller or similar to  the symbol size are not shown.  The red solid line is Kewley  et al. (2006) extreme starburst classification line. The blue-dashed line is the Kauffmann et al. (2003) pure star formation line. The blue solid line is the Seyfert-LINER classification line. The three lines separate galaxies into HII-region-llike, Seyferts, LINERS and composite AGN-HII.}
\label{diag}
\end{figure*}

\begin{table*}
\centering
\caption{Line ratios for the NE hot spot, the Eastern bubble and the circumgalactic gas SW of the QSO2 along PA48. $n_ {\rm e}$ and  $T_ {\rm e}$ could only be measured for the NE hot spot.  $T_ {\rm e}$ has been calculated using the measured  [OIII]4959,5007/[OIII]4363 ratio ($^a$) (which gives a lower limit on $T_ {\rm e}$), and the  reddening corrected ratio ($^b$) based on H$\alpha$/H$\beta$. Notice the very high $T_ {\rm e}$.} 
\begin{tabular}{lllll}
\hline
Ratio &  NE  hot spot  &  Eastern bubble  & Large aperture PA 48\\
	&	PA48 &  Ap. H, PA83 &  SW hot spot + Ap. E + Ap. F \\ \hline
~[OIII]5007/H$\beta$ & 3.9$\pm$0.4  &  7.3$\pm$0.4 & 6.8$\pm$2.3 \\
~[NII]6583/H$\alpha$ & 0.27$\pm$0.03  & $\la$0.12 & 0.30$\pm$0.08 \\
~[SII]6716,6731/H$\alpha$ &  0.58$\pm$0.02  & 0.21$\pm$0.04 &  0.20$\pm$0.05  \\
~[OI]6300/H$\alpha$ & 0.17$\pm$0.03 &   0.09$\pm$0.01  & $\la$0.13 \\ \hline
~H$\gamma$/H$\beta$ & 0.46$\pm$0.06 & 0.44$\pm$0.05 \\
~H$\alpha$/H$\beta$ &  5.3$\pm$0.6  & 2.7$\pm$0.4  & 5.3$\pm$1.8 \\
~HeII4686/H$\beta$ & 0.17$\pm$0.03 \\
~[SII]4068/[SII]6716,6731 & 0.07$\pm$0.02 \\
~[SII]6716/[SII]6731& 1.25$\pm$0.04  & &   \\    
~[OIII]4959,5007/[OIII]4363$^a$ & 44.3$\pm$5.4 &   & \\  
~[OIII]4959,5007/[OIII]4363$^b$ & 30.9$\pm$15.8 &  & \\  \hline
~$n_{\rm e}$ & 184$^{+60}_{-54}$ & & \\
~$T_ {\rm e}^a$  &1.90 $^{+0.15}_{-0.12}$\\  
~$T_ {\rm e}^b$ & 2.40$^{+2.78}_{-0.56}$\\  \hline
\end{tabular}
\label{tab:ratios-circ}
\end{table*}

We show in Fig. \ref{diag} three  standard BPT (Baldwin et al. \citeyear{bal81}) diagnostic diagrams often used to discriminate between different types of galaxies   according to \cite{kew06}: HII-region-type, LINERS, Seyferts and composite AGN-HII. For the ``Beetle'', the different apertures are in general located in the ``Seyfert'' area of the diagram, implying a dominant contribution of an excitation mechanism harder than stellar photoionization. Photoionization by the quasar continuum and shock related mechanisms (Dopita \& Sutherland \citeyear{dop96})
are natural possibilities. This is confirmed by the detection of strong HeII$\lambda$4686 at the location of the NE hot spot with HeII/H$\beta$=0.17$\pm$0.03, which is inconsistent with stellar photoionization and typical of AGN.

We can check whether the AGN can provide sufficient photons to explain the line luminosities of the circumgalactic gas. The total H$\alpha$ luminosity of the NE bubble as measured from the narrow band image is $L_{\rm H\alpha}\sim$1.7$\times$10$^{41}$ erg s$^{-1}$. Correcting  for reddening,  assuming an average H$\alpha$/H$\beta\sim$4.1  (Table \ref{tab:ratios-circ}), an intrinsic $L_{\rm H\beta}\sim$1.3$\times$10$^{41}$ erg s$^{-1}$ is derived.
The ionizing photon luminosity required to power it is $Q_{\rm ion}^{\rm abs}= \frac{L_{\rm H\beta}}{h~\nu_{\rm H\beta}}\frac{\alpha_{\rm B}}{\alpha^{eff}_{H\beta}}=5.1 \times 10^{53}~ \rm s^{-1}$  (Osterbrock \citeyear{ost89})
where  $Q_{\rm ion}^{\rm abs}$ is the ionizing photon luminosity {\it absorbed} by the gas, $\alpha_{\rm B}$=2.5$\times$10$^{-13}$ cm$^3$ s$^{-1}$, $\alpha^{eff}_{H\beta}$=1.6$\times$10$^{-14}$ cm$^3$ s$^{-1}$ for $T_ {\rm e}\sim$2.0$\times$10$^4$ K  as implied by the [OIII] lines and $h \nu_{\rm H\beta}$ is the energy of a H$\beta$ photon.   The NE bubble subtends an angle of $\sim90^{\circ}$  on the plane of the sky around the AGN. Assuming its shape is roughly symmetric  along and perpendicular to the line of sight, the necessary isotropic output from the AGN   is $Q_{\rm ion}^{\rm AGN'}$=3.5$\times$10$^{54}$ s$^{-1}$.

The AGN ionizing luminosity  can be estimated from $L^{\rm AGN}_{\rm ion}\sim 0.35 \times L_{\rm bol} =3.9 \times 10 ^{45}$ erg s$^{-1}$ (Stern et al. \citeyear{ste14}). Assuming $L_ {\nu}\propto \nu^\xi$, with $\xi$=-1.5, gives    $Q^{\rm AGN}_{\rm ion}\sim$5.0$\pm$10$^{55}$ s$^{-1}$. This is thus sufficient to explain the  H$\alpha$ luminosity of the circumgalactic structures (including the SW ones, which are significantly fainter), provided they have  low opacity within their volume.

On the other hand,  the clear correlation between the optical and the radio morphologies suggests that  shocks induced by the interaction with the radio structures contribute to the emission, at least at certain locations. This may also explain the very high $T_ e$ at the NE hot spot location.

\section{Discussion}
\label{Sec:discussion}

We discuss in this section the nature of the large scale outflow mechanism  and the origin of the circumgalactic gas around the ``Beetle''.

\subsection{Giant bubbles inflated by a nuclear AGN wind}
\label{Sec:giantbubbles}

 We inferred in Sect. \ref{Sec:nuclear}, $M_{\rm o}$=3.5$\times$10$^{6}$ M$_{\odot}$, $\dot M _{\rm o}$=2.6 M$_ {\odot}$ yr$^{-1}$,  $\dot E_{\rm o}$=3.2$\times$10$^{41}$ erg s$^{-1}$ (or $\sim$10$^{-5} \times L_{\rm Edd}$) and $\dot p_{\rm o}$=1.0$\times$10$^{34}$ dyne (or $\sim$0.01 $L_{\rm Edd}$/c) for the nuclear outflow.  If this is powered by an AGN wind, both $\dot E_{\rm o}$ and $\dot p_{\rm o}$  are well below the minimum values required   to have a substantial impact on the host galaxy  ($\dot E_{\rm o}\ga$0.05 L$_{\rm Edd}$ and
$\dot p_{\rm o}\ga$20 L$_{\rm Edd}$/c, Zubovas \& King \citeyear{zub12}).     The outflow high density ($n_{\rm e}=$2571$^{+14181}_{-1461}$ cm$^{-2}$) and small size ($R_{\rm o}\la$1 kpc)    are consistent with \cite{fau12}, who showed that such high densities would be sufficient to quench an AGN powered wind prior to reaching radii $\ga$1 kpc.  
Thus, if the nuclear ionized outflow has been generated by an AGN wind, we find no evidence for a significant impact on the host galaxy, in agreement with \cite{vm16} who found this to be a common situation in  QSO2 at $z\la$0.7.

The arguments above are based on the assumption that the AGN driven wind responsible for the nuclear outflow expands in a  high density  medium. 
If  on the contrary it encounters  paths of less resistance  with densities significantly less than the normal volume averaged gas densities, it could advance more freely and  expand to much larger distances  (Nims et al. \citeyear{nims15}, Mukherjee et al. \citeyear{muk16}).  Such scenario has been proposed, for instance, to explain  the large scale radio and optical bubbles associated with the ``Teacup''  QSO2 (Harrison et al. \citeyear{har15}).

 We investigate next whether such scenario could explain the properties (size, velocity, mass) of the circumgalactic optical arcs of the ``Beetle''. 
Following  \cite{nims15} the kinetic energy of the wind can be approximated as $L_{\rm kin}\sim$0.05$\times$$L_{\rm bol}\sim$5$\times$10$^{44}$ erg s$^{-1}$. We use equations 6 and 7 in their paper to calculate the radius $R_{\rm s}$ of the forward shock of the wind and its velocity $v_ {\rm s}$. We use the same fiducial values for $R_ {0}$=100 pc,
$n_{\rm H,0}$=10 cm$^{-3}$, $v_{\rm in}$=0.1 $c$ and the density varying as a power law of index $\beta$=1.0 (see Nims et al. \citeyear{nims15} for details). In 30 Myr, which is a reasonable quasar life time, the wind could create a bubble of radial size $R_{\rm s}\sim$25 kpc as observed for the ``Beetle''. At that time, its  expansion  velocity will be $v_{\rm s}\sim$535 km s$^{-1}$. This is in reasonable agreement with the $V_ {\rm s}$ values measured for the circumgalactic gas (Figs. \ref{fig:kinem}), taking into account that these are line of sight projected velocities.
It thus seems plausible that an AGN wind expanding through paths of low resistance can create the huge bubbles.

 Can this mechanism eject sufficient mass? We have estimated what fraction of the gas dragged  by the nuclear outflow can escape  the gravitational potential of the galaxy. For this, we constrain the gas fraction  that moves with velocity $v$ higher than the escape velocity $v_{esc}$.   $v_{esc}$ for an isothermal sphere at a distance $r$ is given by (Rupke et al. \citeyear{rup02}) $v_ {esc}( r) = \sqrt{2} ~ v_c~[1 + {\rm ln}(r_ {max}/r)]^{1/2}$ where $v_c$ is the circular velocity which scales with the stellar velocity dispersion $\sigma_*$ 
as $v_c = \sqrt{2} \sigma_ {*}$, where $\sigma_*\sim$200 km s$^{-1}$ (Sect. \ref{Sec:object}).  $r_ {max}$ is unknown. We have considered 3 possible truncation values
$r_{max}$=3, 10, 100 kpc. We assume $r=R_ {\rm o}\sim$0.86 kpc, which is the radial size of the outflow (Sect. \ref{Sec:size}).
$v_{\rm esc}$ at $R_{\rm o}$ is in the range $\sim$633-960 km s$^{-1}$ for $r_ {\rm max}$ in the range 3-100 kpc.
Only $\sim$3\% of  the total nuclear $L_{\rm H\beta}$ is   emitted by  gas moving at $v_{\rm esc}>$ 633 km s$^{-1}$. 
Thus, the nuclear outflow can eject only $\sim$0.03$\times M_ {\rm o}$=10$^5$ M$_ {\rm \odot}$ outside the galaxy at a rate of $\sim$0.03$\times \dot M_ {\rm o}$=0.08 M$_{\odot}$ yr$^{-1}$  (Sect. \ref{Sec:mass}).
These are gross upper limits, since higher $v_ {\rm esc}$ are possible.  

The total H$\alpha$ intrinsic (reddening corrected) luminosity of the NE bubble is  $L_{\rm H\alpha}$$\sim$4.0$\times$10$^{41}$ erg s$^{-1}$.  Assuming $n_{\rm e}\sim$184 cm$^{-3}$  as measured  at  the location of the NE hot spot, the total NE bubble mass of the H$\alpha$ emitting gas  is $\sim$2.0$\times$10$^7$ M$_ {\rm \odot}$.  It follows that $>$3.0$\times$10$^8$  yr of continuous mass ejection would have been required to create the NE bubble, which  is longer than any realistic  quasar episode.  The inconsistency gets worse if the mass of the SW arcs are also taken into account. 

Because it will be relevant in Sect. \ref{Sec:origin}, we highlight here that the same conclusion applies if the nuclear outflow was triggered by the small scale radio jet instead of an AGN wind, since the necessary time  of continuous mass ejection is  longer than realistic radio source ages. 

 \subsection{Large scale radio-gas interactions}
\label{Sec:largescale}

The above scenario is also difficult to reconcile with the clear correlation between the morphological, kinematic and ionization properties of the ionized gas and the large scale radio structures (Sect. \ref{Sec:kinem}).
A  more natural scenario is one in which  the radio structures have been originated in  the neighbourhood of the SMBH and have escaped out of the galaxy  boundaries. The  morphological features  (distant  hotspots, a diffuse lobe-like component and  the inner jet-like structure  in the central kpc region) are
well known  constituents of the radio structures in many active galaxies. No convincing alternative explanations (see, for instance, Harrison et al. \citeyear{har15}) for their origin seem likely.

These structures have interacted in their advance with the in-situ (see below) circumgalactic  gas.
  Our preferred scenario is  strongly supported by studies of radio galaxies and radio loud quasars  with clear signs of radio-gas interactions  (e.g. Clark et al. \citeyear{cla98}, Villar Mart\'\i n et al. \citeyear{vm99}, Tadhunter et al. \citeyear{tadh00}, Best et al. \citeyear{best01}, Humphrey et al. \citeyear{hum06}). 
 
 We have seen that (a) the line emission is enhanced at the location of the radio hot spots, especially the NE one, (b) the gas kinematics are turbulent across the spatial range circumscribed by the radio hot spots   (not beyond), (c)   the turbulent gas  shows minima in the ionization level at the location of the radio hot spots, (d) the gas overlapping with the NE hot spot is very hot  $T_ {\rm e}\ga$1.9$\times$10$^4$ K and (e) there is turbulent gas  not only along the radio axis (up to $\sim$26 kpc from the AGN), but also far away from it ($\sim$25 kpc from the radio axis, in the direction perpendicular to it) (Sect. \ref{Sec:ext}). 

 Powerful radio galaxies with radio-gas interactions at different redshifts often show close radio-optical associations, emission line flux enhancement, ionization minima and kinematic turbulence (split lines, broad components) coincident with the radio  structures, anti-correlation between line FWHM and gas ionization level, high $T_ {\rm e}$  have been  seen in such systems. They can all be naturally explained by the interaction between the radio structures and the ambient gas.  The shocks resulting from such interactions create radial outflows that drag, heat and compress the multiphase gaseous medium (Bicknell \citeyear{bic94}, Mukherjee et al. \citeyear{muk16}).

The presence of turbulent gas in the ``Beetle''  up to $\sim$26  kpc  along the radio axis  and $\sim$25 kpc away from it reveals the action of the shocks across a huge volume. This has been observed also in some powerful radio galaxies  (Villar Mart\'\i n et al. \citeyear{vm99}, Tadhunter et al. \citeyear{tadh00}). Turbulent kinematics up to $\sim$6 kpc from the AGN  have also been reported for the ``Teacup'' radio quiet QSO2 (Keel et al. \citeyear{keel17}, Ramos Almeida et al. \citeyear{ram17}). Although a scenario where the outflows have been triggered by an AGN wind (rather than by the radio structures) cannot be rejected, it is possible that a similar mechanism as the one inferred for the ``Beetle'' is at work (Harrison et al. \citeyear{har15}).

 \cite{gas12}  have shown that a radio source of similar power as the ``Beetle''   can  create  a wealth of  observable features around ellipticals in poor environments, such as X-ray cavities surrounded  by weak shocks, large buoyant bubbles, extended filamentary multiphase gas features (including cold gas with $T_e\la$20,000 K gas) and subsonic chaotic turbulence.  The effects can be visible up to  several kpc from the SMBH.   Whether this can occur up to $\sim$26 kpc is not clear  from the point of view of the simulations.

Powerful radio galaxies with large volumes of gas affected by radio induced outflows  (Villar Mart\'\i n et al. \citeyear{vm99}, Tadhunter et al. \citeyear{tadh00}), show wide area radio features (e.g. radio lobes)  overlapping with the optical bubbles or bow shocks. This is not the case in the ``Beetle'', but this is not an inconsistency of our proposed scenario.  Although a relativistic jet may appear to be collimated in radio,  it will in fact launch a bubble defined by the forward shock (Bicknell  \citeyear{bic94}, Mukherkee et al. \citeyear{muk16}).  Hence although the radio contours may trace predominantly the central axis, where the electrons are relativistic and  emit intense  synchrotron radiation, the outer shock will be larger,  more dilute and  difficult to detect. This may explain why broad radio lobes overlapping with the H$\alpha$ bubble edges have not been detected in the ``Beetle''. A deeper radio scan may reveal them.

 We can apply the methodology of \cite{nims15}   (see Sect. \ref{Sec:giantbubbles}) to investigate whether a radio source of the same mechanical power as that of the ``Beetle'' can inflate the huge observed bubbles. The equations   were originally derived to model the expansion of a homogeneous spherically symmetric wind. While strong fast jets produce asymmetric bubbles with the jet head progressing much faster, weak jets (as that in the ``Beetle'') evolve more spherically (Mukherjee et al. \citeyear{muk16}). We thus  consider  the   equations valid as well in the scenario where the outflow has been induced by the  radio  source. The mechanical power of the radio source can be calculated   as $P_{\rm jet} \sim  5.8\times$10$^{43}  (P_{\rm radio}/10^{40})^{0.70}$ erg s$^{-1}$  (Cavagnolo et al. \citeyear{cav10}).
	For the ``Beetle'', we infer $P_{\rm radio}$ = 9.1$^{+45.3}_{-5.9} \times 10^{40}$ erg s$^{-1}$ using the radio power at 1.4 GHz, integrating between 10 MHz and 30 GHz, considering the range of expected spectral index $\alpha$=0.094$^{+0.76}_{-1.01}$ and $F_ {\nu} \propto \nu^{\alpha}$. Hence, $P_{\rm jet}= 2.7^{+6.8}_{-1.4} \times 10^{44}$  erg s$^{-1}$.  In 35 Myr, which is a reasonable radio source age (Parma et al. \citeyear{par99}), the jet could therefore create a bubble of radial size $R_{\rm s}\sim$25 kpc. At that time, its  expansion  velocity will be $v_{\rm s}\sim$442 km s$^{-1}$.

  \subsection{The origin of the circumgalactic gas}
\label{Sec:origin}

As we saw in Sect. \ref{Sec:giantbubbles}, it is unlikely that the circumgalactic features consist of gas ejected from the galaxy by the nuclear outflow, independently  of whether this has been triggered by and AGN wind or the small scale radio jet.  It appears more likely that they consist of in-situ gas, maybe tidal remnants redistributed during  galactic interactions across  $\sim$71 kpc (Sect. \ref{Sec:extension}), as observed.  Such events are expected to populate the CGM with tidal debris. The circumgalatic  gas around the ``Beetle'' has been shaped and rendered visible due to the illumination by the QSO and the action of the radio induced outflows. The possible detection of an infalling circumgalactic  reservoir of gas (Sect. \ref{Sec:kinem}) could point
 to a scenario  in which part of  the CGM  is raining back, falling towards the gravitational centre probably defined by  the QSO2 host.

\subsection{Is the ``Beetle'' related to high excitation radio galaxies?} 
\label{Sec:herg}

The novelty of our results lies on the fact that  the ``Beetle'' is radio quiet.  As explained  is Sect. \ref{Sec:intro} radio-induced feedback has been rarely considered as potentially relevant in radio quiet luminous AGN, especially  across large spatial scales ($\gg$ several kpc).   In this section we analyze in more depth the implications of our results, by comparing the ``Beetle'' with radio galaxies.

Radio galaxies can be classified into two distinct classes according to the optical emission line ratios and the [OIII]$\lambda$5007 equivalent width:  high (HERG) and low excitation (LERG) radio sources, which  show profound differences (e.g. Laing et al. \citeyear{lai94}, Tadhunter et al. \citeyear{tadh98}, Best \& Heckman \citeyear{best12}). Contrary to LERG, HERG have high luminosity accretion disks (+ X ray corona), bright line emission, NIR/sub-mm evidence of dusty obscuring torus and orientation-dependent properties. Some fundamental differences have been interpreted in terms of differences on the underlying accretion modes. In HERG, accretion at high Eddington ratios proceeds via a geometrically thin, optically thin  accretion disk with a high radiative efficiency, while LERG are powered by advection-dominated accretion flows (ADAF, Narayan \& Yi \citeyear{nar94}) with low radiative efficiency. \cite{har07} show that LERG are predominantly low-power Fanaroff-Riley class I (FR-I) sources, while the HERG population consists of both FR-I and the more powerful FR-II sources, whose relativistic jets end in bright hot-spots. The dominant feedback mechanism is also thought to be clearly distinct between LERG and HERG. LERG are dominated by radio-mode feedback, where the bulk of the energy is ejected in kinetic form through radio jets (e.g., Croton et al. \citeyear{cro06}) and efficiently coupled to the galaxies' gaseous environment (Best et al. \citeyear{best06}). The radiative (or quasar) mode  is thought to dominate in HERG. 

\begin{figure*}
\includegraphics{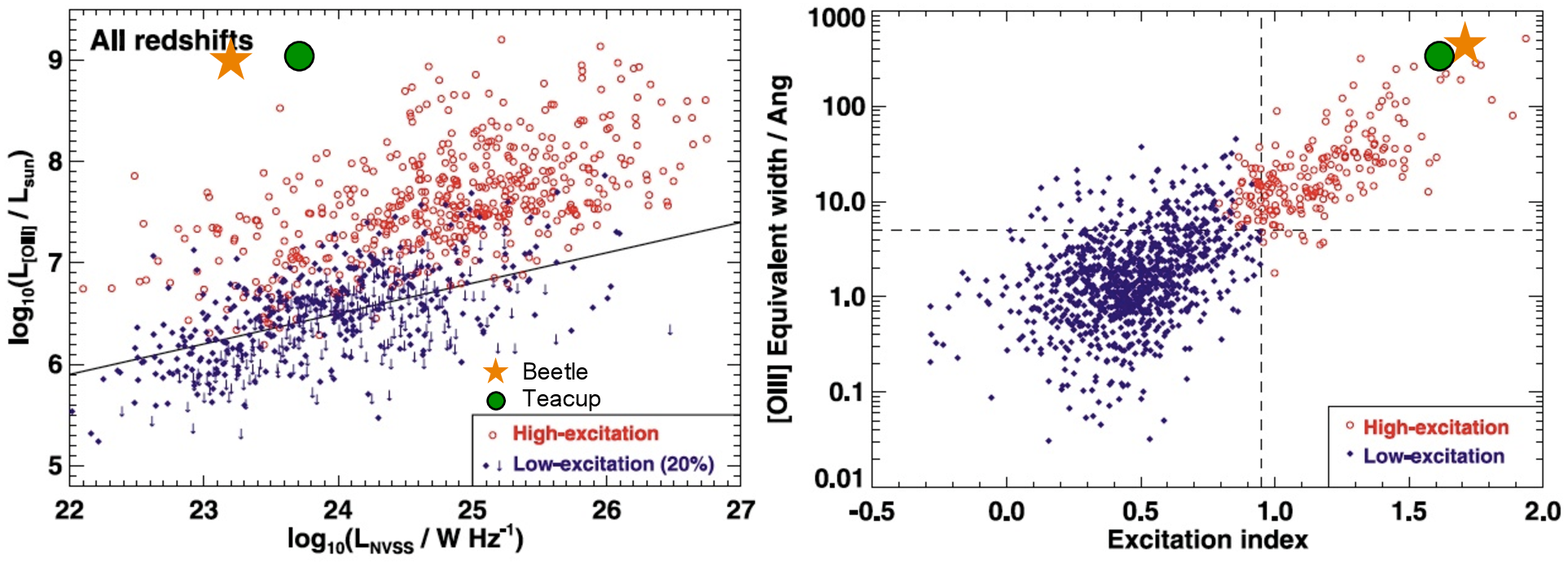}
\vspace{2.2in}
\caption{Figures 1 and 2 from Best \& Heckman (2012) showing the location of the ``Beetle'' (beige star) in comparison with HERG and LERG. The ``Teacup'' (Harrison et al. 2015) is also shown (green circle). From the optical spectroscopic point of view both objects are  HERG with rather extreme values of both [OIII]$\lambda$5007 equivalent width and excitation index (right). Their relatively low radio luminosity, on the other hand,  places them well above  the sequence defined by the the radio galaxy sample in the left diagram.}
\label{herg}
\end{figure*}

How does the ``Beetle''  fit into this picture? Based on its optical spectroscopic properties,  the ``Beetle''   is indistinguishable from HERG (Fig. \ref{herg}).  It has a very high excitation index (Best \& Heckman \citeyear{best12}) $EI$ = ${\rm log_{10}([OIII]/H\beta) -  \frac{1}{3} [log_{10}([NII]/H\alpha})$ ${\rm + log_{10} ([SII]/H\alpha)+ log_{10}([OI]/H\alpha)]=1.70}$,  significantly higher than the limiting value 0.95 proposed by \cite{best12} to separate LERG and HERG. Indeed, it is at the high end of EW$_{\rm [OIII]\lambda 5007}$=430 \AA\  (rest frame value) and $EI$ values spanned by the radio galaxy sample studied by these authors. It has an Eddington ratio  $\lambda=\frac{\rm L_{\rm bol}}{\rm L_{\rm Edd}}$=0.33 (Sect. \ref{Sec:object}). The transition between the radiatively efficient  accretion regime and the radiatively inefficient accretion regime typically occurs around $\lambda\sim$0.01, i.e., 1\% of the Eddington rate, below which, the kinetic output dominates over the radiative output. The ``Beetle'' is accreting at a very high rate. In fact, few radio galaxies  accrete at  such high rates (Best \&  Heckman \citeyear{best12}, Ishibisahi et al. \citeyear{ish14},  Sikora et al. \citeyear{sik13}). 

The rest frame radio power of the ``Beetle''  (log(P$_ {\rm 1.4GHz}$)=30.2 in erg s$^{-1}$ Hz$^{-1}$), although modest, is within the range measured in radio galaxies. A critical difference is that its  very high bolometric luminosity makes it a radio quiet system.   HERG of similar $L_ {\rm [OIII]}$ show $\ga$several$\times$100 times  more powerful radio sources (Fig. \ref{herg},    
 left). Following \cite{ish14}  we define the radio loudness parameter $R'=\frac{\rm L_{\rm 1.4GHz}}{\rm L_{\rm [OIII]}}$, where L$_{\rm 1.4GHz}$ is the rest frame luminosity at 1.4GHz. For the ``Beetle''  log($R'$)=-2.78  below the range spanned by radio galaxies (both HERG and LERG) which show log($R'$)$\ga$-2.5 (Ishibashi et al. \citeyear{ish14}, Best \& Heckman \citeyear{best12}). 

 LERG and HERG show an anti-correlation between log($R'$) and  log($\lambda'$) where the new accretion rate $\lambda'$=3500$\times \frac{L_ {\rm [OIII]}}{L_{\rm Edd}}$=0.42 is  calculated in coherence with \cite{ish14}\footnote{The authors assume $L_{\rm bol}=3500 \times L_ {\rm [OIII]}$ following  \cite{hec04}}. This is likely  a result of either an enhancement of the radio jet emission at low accretion rate and/or a decrease in the optical emission due to the decline of the radiative efficiency in the radiatively inefficient accretion flow (RIAF) mode. The best fit is different for HERG  and LERG (Buttiglione et al. \citeyear{but10}, Ishibashi et al. \citeyear{ish14}), being much steeper for the former. For the ``Beetle'', log($\lambda'$)=-0.37 and log($R'$)=-3.28. All uncertainties considered, including the scatter of the correlations, these values   are consistent with  the HERG fit (expected log($R'$)=-2.6)  and inconsistent with the LERG fit (expected 
 log($R'$)=-0.65).

Interestingly, the ``Teacup'' is very similar to the ``Beetle''  in numerous aspects. It is also a radio quiet luminous QSO2 hosted by a bulge dominated system. It shows prominent shells indicative of previous merger activity. The axis of the radio and optical bubbles ($\sim$10-12 kpc in size) runs roughly perpendicular to the main shells axis, as in the ``Beetle'' (Keel et al. \citeyear{keel12}, Harrison et al. \citeyear{har15}, Ramos Almeida et al. \citeyear{ram17})  and the large scale outflow may also have been triggered by the radio structures (Harrison et al. \citeyear{har15}). We have used the following information published in the literature: the optical nuclear line ratios (Keel et al. \citeyear{keel12}), $L_{\rm [OIII]}$=3.8$\times$10$^{43}$ erg s$^{-1}$, log(P$_{\rm 1.4GHz}$)=30.2 in erg s$^{-1}$ Hz$^{-1}$ (Harrison et al. \citeyear{har15}),  log($M_ {\rm BH}$)=8.4, indirectly inferred from the bulge stellar mass log($M_{\rm sph}/M_{\rm \odot}$)=10.8 (Vizier catalogue, based on Simard et al. \citeyear{sim11}). We obtain $EI$=1.52, log($\lambda'$)=-0.10   and log($R'$)=-3.16, i.e. very high excitation index,  very high accretion rate  and very low radio loudness parameter, demonstrative of its radio quietness (Fig. \ref{herg}). As for the ``Beetle'', these values are consistent with the extrapolation of the  log($R'$) vs.  log($\lambda'$)  correlation of HERG  into the radio quiet regime and inconsistent with the LERG correlation. 

The properties of the ``Beetle'' and the ``Teacup"   suggest that they are high excitation radio quiet QSO2 which are similar to the population of HERG,  extended into the regime of very high accretion efficiencies and modest radio luminosities, i.e.,  the radio quiet regime. This is consistent with \cite{lacy01} who proposed that there is a continuous variation of radio luminosity in quasars and there is no evidence of a ``switch'' at some set of critical parameter values that turns on powerful radio jets.

Based on the very high accretion rate and the radio quietness (or very low $R'$) radiative feedback would be expected to play a dominant role in both objects. We have seen, however, that although this mechanism cannot be discarded in the central region near the AGN ($\la$1 kpc), radio induced feedback dominates in the ``Beetle'' (maybe also the ``Teacup'')  across a huge volume of many 10s of kpc in size, reaching regions well outside the galaxy boundary and into the circumgalactic medium. This opens the possibility that radio mode feedback may contribute to the re-heating of the cooling gas in massive galaxies not only in radio loud systems (Best et al. \citeyear{best06}), but also in at least some radio quiet AGN.

\section{Summary and conclusions}
\label{Sec:conclusions} 

We have presented a detailed study of the ``Beetle'', a radio quiet QSO2   at $z=$0.103, based  on GTC optical images and  long slit spectroscopy as well as VLA radio maps. 

\begin{itemize}

\item The ``Beetle''  shows  clear morphological signs of a tidal interaction (shells, filaments), probably with a star forming companion galaxy located at $\sim$20 kpc.   The system is associated with  a spectacular set of circumgalactic ionized knots and three arcs  reminiscent of optical bow shocks or bubbles, with a main axis aligned with the radio axis. They are located at at 19, 25 and 26 kpc from the AGN.  The   maximum total extension of the ionized gas   is measured along the radio axis ($\sim$71 kpc), with maximum extension from the QSO2  $\sim$41 kpc. These circumgalactic features are ionized by shocks and/or the quasar continuum.

\item The radio map has revealed the existence of a previously unknown  extended radio source, with a  central jet $\sim$4.3 kpc long and two hot spots separated by $\sim$46 kpc, well outside the galaxy boundaries.  The inner  jet and the outer radio axis form a 19$\degr$ angle, which would indicate a substantial precession of the radio jet  or its bending as it propagates outward.

\item The morphological, ionization and kinematic properties of the large scale ionized gas are correlated with the radio structures.   We have seen that (a) the line emission is enhanced at the location of the radio hot spots, especially the NE one, (b) gas kinematic  turbulence (FWHM$\sim$380-470 km s$^{-1}$)   is detected across the spatial range circumscribed by the radio hot spots   (not beyond), (c)   the turbulent gas  shows minima in the ionization level at the location of the radio hot spots, (d) the gas overlapping with the NE hot spot is very hot  $T_ {\rm e}\ga$1.9$\times$10$^4$ K and (e) there is turbulent gas  not only along the radio axis (up to $\sim$26 kpc from the AGN), but also far away from it ($\sim$25 kpc from the radio axis, in the direction perpendicular to it). In addition, a reservoir of more quiescent gas has also been detected whose kinematic pattern suggests infall.

\item  We  propose the following scenario: the interaction between the QSO2 host and the nearby companion galaxy has redistributed galactic material (stars and gas) across $\sim$70 kpc, at opposite sides of both  galaxies. Part of the gas is in the process of falling  towards the gravitational centre, probably defined by the QSO2 host. The radio structures have escaped from the vicinity of the active nucleus out into the circumgalactic medium. They interact with the gas located in their path. Shocks drag, heat and compress the multiphase medium within a huge volume
 enhancing the line emission, perturbing the kinematics and changing the physical properties up to $\sim$25 kpc along the radio axis and perpendicular to it. Our preferred scenario is  strongly supported by studies of radio galaxies and radio loud quasars at different redshifts with clear signs of radio-gas interactions.

\item Specific 3D hydrodynamic simulations of interactions between AGN  jets and the multiphase ISM   adapted to the ``Beetle''  would be of great value  to quantify accurately how radio induced feedback is affecting its large scale environment in terms of energy injection, heating and mass ejection. Ultimately, this would show whether the radio induced feedback can have a significant impact on the evolution of the QSO2 host by means of quenching star formation and/or ejecting a significant fraction of the galaxy ISM.

\item Large scale effects ($\gg$ several kpc) of radio induced feedback  in radio quiet AGN are usually not expected.  
Our results demonstrate that this idea needs to be reconsidered. We have found that jets of modest power can remain strong enough in radio quiet objects 
for length scales of 10s of kpc to interact with the galactic and circumgalactic  environment and provide a feedback mechanism 
that can affect huge volumes in and around galaxies.

\item  The radio and optical properties of the ``Beetle'' imply a very high accretion index, very high accretion rate and a very low radio loudness parameter compared with high excitation radio galaxies (HERG).  We propose that the ``Beetle''  (possibly also the famous ``Teacup" radio quiet QSO2 at $z\sim$0.085,  Keel et al. \citeyear{keel12}, Gagne et al. \citeyear{gag14})  are high excitation radio quiet QSO2 which are similar to the population of HERG,   extended into the  radio quiet regime of objects with  high bolometric luminosities and low or moderate radio luminosities. The ``Beetle'' demonstrates that jets of modest power can be the dominant feedback mechanism acting across huge volumes in highly accreting luminous radio quiet AGN, where radiative (or quasar) mode feedback would be expected to dominate.

\end{itemize}

\section*{Acknowledgments}

Partly based on observations made with the GTC  telescope, in the Spanish Observatorio del Roque de los Muchachos of the Instituto de Astrof\'\i sica de Canarias, under Director's Discretionary Time.
 We  thank the GTC staff for their support with the optical imaging and spectroscopic observations and the VLA staff for executing the radio observations. Thanks to Gustaaf van Moorsel for his valuable advice about the radio observations.
 
 MVM, AC and BE acknowledge  support  from the Spanish Ministerio de Econom\'\i a y Competitividad through the grants AYA2015-64346-C2-2-P and AYA2012-32295. AH acknowledges Funda\c{c}\~{a}o para a Ci\^{e}ncia e a
Tecnologia (FCT) support through UID/FIS/04434/2013, and through
project FCOMP-01-0124-FEDER-029170 (Reference FCT
PTDC/FIS-AST/3214/2012) funded by FCT-MEC (PIDDAC) and FEDER
(COMPETE), in addition to FP7 project PIRSES-GA-2013-612701. AH also
acknowledges a Marie Curie Fellowship co-funded by the FP7 and the FCT
(DFRH/WIIA/57/2011) and FP7 / FCT Complementary Support grant
SFRH/BI/52155/2013. BE acknowledges support from 
the European Union 7th Framework Programme (FP7-PEOPLE-2013-IEF)
under grant 624351.  CRA acknowledges the Ram\'on y Cajal Program of the Spanish Ministry of Economy and Competitiveness through project
RYC-2014-15779 and the Spanish Plan Nacional de Astronom\' ia y Astrofis\' ica under grant AYA2016-76682-C3-2-P.

 This research has made use of: 1) the VizieR catalogue access tool, CDS,
 Strasbourg, France. The original description of the VizieR service was
 published in Ochsenbein et al. A\&AS, 143, 23;   2)  the NASA/IPAC circumgalactic Database (NED) which is operated by the Jet Propulsion Laboratory, California Institute of Technology, under contract with the National Aeronautics and Space Administration; 3) data from Sloan Digital Sky Survey. Funding for the SDSS and SDSS-II has been provided by the Alfred P. Sloan Foundation, the Participating Institutions, the National Science Foundation, the U.S. Department of Energy, the National Aeronautics and Space Administration, the Japanese Monbukagakusho, the Max Planck Society, and the Higher Education Funding Council for England. The SDSS Web Site is http://www.sdss.org/.

\end{document}